\documentclass[longauth]{aa}

\usepackage{txfonts}
\usepackage{graphicx}
\usepackage{natbib}
\bibpunct{(}{)}{;}{a}{}{,}

\begin{document}

\title{The ALHAMBRA survey\thanks{Based on observations collected at the German-Spanish Astronomical Center, Calar Alto (CAHA), jointly operated by the Max-Planck-Institut f\"ur Astronomie (MPIA) at Heidelberg and the Instituto de Astrof\'{\i}sica de Andaluc\'{\i}a (CSIC)}: $B$-band luminosity function of quiescent and star-forming galaxies at $0.2 \leq z < 1$ by PDF analysis}

\author{C.~L\'opez-Sanjuan\inst{\ref{a1}} \and
E.~Tempel\inst{\ref{c3},\ref{c1}} \and
N.~Ben\'itez\inst{\ref{a2}} \and
A.~Molino\inst{\ref{b2},\ref{a2}} \and
K.~Viironen\inst{\ref{a1}} \and
L.~A.~D{\'{\i}}az-Garc{\'{\i}}a\inst{\ref{a1}} \and
A.~Fern\'andez-Soto\inst{\ref{a12},\ref{a5}} \and 
W.~A.~Santos\inst{\ref{b2}} \and
J.~Varela\inst{\ref{a1}} \and
A.~J.~Cenarro\inst{\ref{a1}} \and
M.~Moles\inst{\ref{a1},\ref{a2}} \and
P.~Arnalte-Mur\inst{\ref{a3},\ref{a4}} \and
B.~Ascaso\inst{\ref{b3}} \and
A.~D.~Montero-Dorta\inst{\ref{b4}} \and
M.~Povi\'c\inst{\ref{a2}} \and
V.~J.~Mart\'inez\inst{\ref{a3},\ref{a4},\ref{a5}} \and
L.~Nieves-Seoane\inst{\ref{a12},\ref{a3},\ref{a4}} \and
M.~Stefanon\inst{\ref{c2}} \and
Ll.~Hurtado-Gil\inst{\ref{a3}} \and
I.~M\'arquez\inst{\ref{a2}} \and
J.~Perea\inst{\ref{a2}} \and
J.~A.~L.~Aguerri\inst{\ref{a10},\ref{a11}} \and
E.~Alfaro\inst{\ref{a2}} \and
T.~Aparicio-Villegas\inst{\ref{b1},\ref{a2}} \and
T.~Broadhurst\inst{\ref{a6},\ref{a7}} \and
J.~Cabrera-Ca\~no\inst{\ref{a8}} \and
F.~J.~Castander\inst{\ref{a9}} \and
J.~Cepa\inst{\ref{a10},\ref{a11}} \and
M.~Cervi\~no\inst{\ref{a2},\ref{a10},\ref{a11}} \and
D.~Crist\'obal-Hornillos\inst{\ref{a1}} \and
R.~M.~Gonz\'alez~Delgado\inst{\ref{a2}} \and
C.~Husillos\inst{\ref{a2}} \and
L.~Infante\inst{\ref{a13}} \and  
J.~Masegosa\inst{\ref{a2}} \and
A.~del~Olmo\inst{\ref{a2}} \and
F.~Prada\inst{\ref{a2},\ref{a14},\ref{a15}} \and 
J.~M.~Quintana\inst{\ref{a2}}
}

\institute{Centro de Estudios de F\'isica del Cosmos de Arag\'on (CEFCA), Unidad Asociada al CSIC, 
            Plaza San Juan 1, 44001 Teruel, Spain\\\email{clsj@cefca.es}\label{a1} 
            \and
Leibniz-Institut für Astrophysik Potsdam (AIP), An der Sternwarte 16, 14482 Potsdam, Germany\label{c3} 
            \and
Tartu Observatory, Observatooriumi 1, 61602 T\~oravere, Estonia\label{c1} 
	    \and
IAA-CSIC, Glorieta de la Astronom\'ia s/n, 18008 Granada, Spain\label{a2} 
	    \and
Instituto de Astronom{\'{\i}}a, Geof{\'{\i}}sica e Ci\'encias Atmosf\'ericas, 
              Universidade de S{\~{a}}o Paulo, 05508-090 S{\~{a}}o Paulo, Brazil\label{b2}
	    \and
Instituto de F\'isica de Cantabria (CSIC-UC), 39005 Santander, Spain\label{a12}
            \and
Unidad Asociada Observatorio Astron\'omico (IFCA-UV), 
            46980 Paterna, Spain\label{a5}
	    \and
Observatori Astron\`omic, Universitat de Val\`encia, 
            C/ Catedr\`atic Jos\'e Beltr\'an 2, 46980 Paterna, Spain\label{a3}
	    \and
Departament d'Astronomia i Astrof\'isica, 
            Universitat de Val\`encia, 46100 Burjassot, Spain\label{a4}
            \and
APC, AstroParticule et Cosmologie, Universit\'e Paris Diderot, CNRS/IN2P3, CEA/lrfu, Observatoire de Paris, Sorbonne Paris Cit\'e, 10, rue Alice Domon et L\'eonie Duquet, 75205 Paris Cedex 13, France\label{b3}
	    \and
Department of Physics and Astronomy, University of Utah, 115 South 1400 East, Salt Lake City, UT 84112, USA\label{b4}
            \and
Leiden Observatory, Leiden University, P.O. Box 9513, 2300 RA Leiden, The Netherlands\label{c2}
 	    \and
Instituto de Astrof\'isica de Canarias, V\'ia L\'actea s/n, 38200 
              La Laguna, Spain\label{a10}
            \and
Departamento de Astrof\'isica, Facultad de F\'isica, 
              Universidad de La Laguna, 38206 La Laguna, Spain\label{a11}
            \and
Observat\'orio Nacional-MCT, Rua Jos\'e Cristino 77, CEP 20921-400, 
              Rio de Janeiro-RJ, Brazil\label{b1}
            \and
Department of Theoretical Physics, 
            University of the Basque Country UPV/EHU, 48080 Bilbao, Spain\label{a6}
            \and
IKERBASQUE, Basque Foundation for Science, 48013 Bilbao, Spain\label{a7}
            \and
Departamento de F\'isica At\'omica, Molecular y Nuclear, 
             Facultad de F\'isica, Universidad de Sevilla, 41012 Sevilla, Spain\label{a8}
            \and
Institut de Ci\`encies de l'Espai (IEEC-CSIC), Facultat de Ci\`encies, 
             Campus UAB, 08193 Bellaterra, Spain\label{a9}
            \and
Departamento de Astronom\'ia, Pontificia Universidad Cat\'olica. 
              782-0436 Santiago, Chile\label{a13}
            \and
Instituto de F\'{\i}sica Te\'orica (UAM/CSIC), Universidad Aut\'onoma 
              de Madrid, Cantoblanco, 28049 Madrid, Spain \label{a14}
            \and
Campus of International Excellence UAM+CSIC, Cantoblanco, 
               28049 Madrid, Spain \label{a15}
}

\date{Received 11 August 2016 / Accepted 21 November 2016}

\abstract
{}
{Our goal is to study the evolution of the $B$-band luminosity function (LF) since $z \sim 1$ using ALHAMBRA data.}
{We used the photometric redshift and the $I$-band selection magnitude probability distribution functions (PDFs) of those ALHAMBRA galaxies with $I \leq 24$ mag to compute the posterior LF. We statistically studied quiescent and star-forming galaxies using the template information encoded in the PDFs. The LF covariance matrix in redshift - magnitude - galaxy type space was computed, including the cosmic variance. That was estimated from the intrinsic dispersion of the LF measurements in the 48 ALHAMBRA sub-fields. The uncertainty due to the photometric redshift prior is also included in our analysis.}
{We modelled the LF with a redshift-dependent Schechter function affected by the same selection effects than the data. The measured ALHAMBRA LF at $0.2 \leq z < 1$ and the evolving Schechter parameters both for quiescent and star-forming galaxies agree with previous results in the literature. The estimated redshift evolution of $M_{B}^* \propto Qz$ is $Q_{\rm SF} = -1.03 \pm 0.08$ and $Q_{\rm Q} = -0.80 \pm 0.08$, and of $\log_{10}\phi^{*} \propto Pz$ is $P_{\rm SF} = -0.01 \pm 0.03$ and $P_{\rm Q} = -0.41 \pm 0.05$. The measured faint-end slopes are $\alpha_{\rm SF} = -1.29 \pm 0.02$ and $\alpha_{\rm Q} = -0.53 \pm 0.04$. We find a significant population of faint quiescent galaxies with $M_B \gtrsim -18$, modelled by a second Schechter function with slope $\beta = -1.31 \pm 0.11$.}
{We present a robust methodology to compute LFs using multi-filter photometric data. The application to ALHAMBRA shows a factor $2.55\pm0.14$ decrease in the luminosity density $j_B$ of star-forming galaxies, and a factor $1.25\pm0.16$ increase in the $j_B$ of quiescent ones since $z = 1$, confirming the continuous build-up of the quiescent population with cosmic time. The contribution of the faint quiescent population to $j_B$ increases from $3$\% at $z = 1$ to 6\% at $z = 0$. The developed methodology will be applied to future multi-filter surveys such as J-PAS.}


\titlerunning{The ALHAMBRA $B$-band luminosity function at $0.2 \leq z < 1$ by PDF analysis.}

\authorrunning{L\'opez-Sanjuan et al.}

\maketitle

\section{Introduction}\label{intro}
The greatest advances in the galaxy formation and evolution field in the last decade have been possible thanks to systematic extragalactic surveys, both photometric and spectroscopic. However, even if the general trends in galaxy properties (luminosity and mass function, star formation rate, metallicity, morphology and structure, etc.) and their redshift evolution are qualitatively established, the particular physical processes causing these trends and their relative role in galaxy formation are still under debate. To unveil such physical processes, we must quantify with exquisite details not only the distribution of galaxy properties, but also their intrinsic (physical) dispersions and possible correlations.

In the next decade, large-area photometric surveys such as J-PAS\footnote{\url{j-pas.org}} (Javalambre – Physics of the accelerating universe Astrophysical Survey; \citealt{jpas}), Euclid \citep{euclid}, and LSST (Large Synoptic Survey Telescope; \citealt{lsst}), will provide unprecedented statistical power to derive precision galaxy distributions and eventually disentangle the physics behind them. The multi-filter photometric survey J-PAS will observe 8500 deg$^2$ of the northern hemisphere with 56 narrow-band filters ($\sim$145\AA), providing $R \sim 50$ photo-spectra of about 200 million sources, leading to a photometric redshift precision of $\sim$1000 km s$^{-1}$, and allowing emission line and stellar continuum measurements.

However, the statistical J-PAS strength is also its main challenge: with statistical uncertainties being no longer a problem, the systematics in the analysis techniques will dominate the final error budget. Usual photometric techniques are prone to known biases \citep{sheth10} and the J-PAS photo-spectra resolution is too low to successfully apply spectroscopic tools, so new and more well-suited methodologies are mandatory to extract robust, unbiased, and accurate J-PAS galaxy distributions for the next decade astrophysics. There are several ways to attack this problem: we can deconvolve the observed photometric distributions \citep[e.g.][]{sheth10,rossi10,taylor15,monterodorta16lf}, use the posterior probability distribution functions (PDFs) of the parameters \citep[e.g.][]{sheth10}, or apply sophisticated statistical methods \citep[e.g.][]{lake16}.

To address the J-PAS technical challenges, the PROFUSE\footnote{\url{profuse.cefca.es}} project uses PRObability Functions for Unbiased Statistical Estimations in multi-filter surveys, developing novel techniques based on posterior PDFs to analyse photometric data. Even if the posterior PDFs are recognised as the right way to deal with photometric redshifts \citep[e.g.][]{soto02,cunha09,wittman09,myers09,schmidt13,carrasco14a,asorey16} and Bayesian inference is widely used to estimate galaxy properties, current distribution estimators assume galaxies with a fixed $z$, luminosity, stellar mass, amongst others. However, given the probabilistic nature of the photometric redshifts, any galaxy property becomes also probabilistic and thus the posterior PDFs must be tracked along the analysis process to ensure unbiased galaxy distributions.

The luminosity function, that is, the number of galaxies per unit volume and magnitude interval, is a powerful tool to study galaxy evolution, and it is estimated by virtually any extragalactic survey \citep[see][for a recent review]{Johnston11}. It provided the first insights about the emergence of the red population and the star-formation quenching of blue galaxies since $z \sim 1$ \citep{bell04,faber07}. Because of its fundamental significance, in this paper we present the PROFUSE estimation of the $B-$band luminosity function using the multi-filter ALHAMBRA\footnote{\url{www.alhambrasurvey.com}} (Advanced, Large, Homogeneous Area, Medium-Band Redshift Astronomical) survey \citep{alhambra}. The rest-frame $B-$band is well covered by extragalactic optical surveys, allowing the study of non-extrapolated luminosities up to $z \sim 1$ \citep[see][for a recent compilation of $B-$band luminosity functions]{beare15}.

The PROFUSE estimator of the luminosity function has important advantages with respect to previous ones. Our new method provides a posterior luminosity function, $\Phi\,(z,M_B)$, and (i) naturally accounts for $z$ and $M_B$ uncertainties, (ii) ensures 100\% completeness because it works with intrinsic magnitudes instead than with observed ones, (iii) robustly deals with spectral type selections because we can statistically decompose the luminosity function on quiescent and star-forming populations, and (iv) provides a reliable covariance matrix in redshift--magnitude--galaxy type space. Moreover, instead of studying the luminosity function in redshift slices, we created a model in $z - M_B$ that is affected by the same selection as the data, avoiding volume incompleteness, using all the available galaxies to infer the model parameters, and minimising the impact of cosmic variance over the redshift.

This paper is organised as follows. In Sect.~\ref{data} we present the ALHAMBRA survey, and its photometric redshifts and posterior distributions. We develop the methodology to measure the luminosity function by PDF analysis in Sect.~\ref{lfal}. We present the estimated ALHAMBRA luminosity function of both star-forming and quiescent galaxies in Sect.~\ref{results}, and discuss our results in Sect.~\ref{discussion}. Finally, we summarise our work and present our conclusions in Sect.~\ref{conclusion}. Throughout this paper we use a standard cosmology with $\Omega_{\rm m} = 0.3$, $\Omega_{\Lambda} = 0.7$, $\Omega_{\rm k} = 0$, $H_{0}= 100h$ km s$^{-1}$ Mpc$^{-1}$, and $h = 0.7$. The results from previous studies were converted to our cosmology. Magnitudes are given in the AB system \citep{oke83}. For clarity, scalars are represented as $\Phi$, vectors as $\vec{\Phi}$, and tensors as $\tens{\Phi}$.

\begin{table*}
\caption{First data release ALHAMBRA survey fields.}
\label{alhambra_fields_tab}
\begin{center}
\begin{tabular}{lcccc}
\hline\hline\noalign{\smallskip}
Field      &    Overlapping     &    RA    &    DEC     &    sub-fields / area \\
name       &      survey        &   (J2000) & (J2000)    &   (no. / deg$^2$)\\
\noalign{\smallskip}
\hline
\noalign{\smallskip}
ALHAMBRA-2  &  DEEP2    \citep{deep2}   & 02 28 32.0    & +00 47 00   &  8 / 0.377    \\
ALHAMBRA-3  &  SDSS     \citep{sdssdr8} & 09 16 20.0    & +46 02 20   &  8 / 0.404    \\
ALHAMBRA-4  &  COSMOS   \citep{cosmos}  & 10 00 00.0    & +02 05 11   &  4 / 0.203    \\
ALHAMBRA-5  &  GOODS-N  \citep{goods}   & 12 35 00.0    & +61 57 00   &  4 / 0.216    \\
ALHAMBRA-6  &  AEGIS    \citep{aegis}   & 14 16 38.0    & +52 24 50   &  8 / 0.400    \\
ALHAMBRA-7  &  ELAIS-N1 \citep{elais}   & 16 12 10.0    & +54 30 15   &  8 / 0.406    \\
ALHAMBRA-8  &  SDSS     \citep{sdssdr8} & 23 45 50.0    & +15 35 05   &  8 / 0.375    \\
Total       &           &               &             & 48 / 2.381\\
\noalign{\smallskip}
\hline
\end{tabular}
\end{center}
\end{table*}

\section{ALHAMBRA survey}\label{data}
The ALHAMBRA survey provides a deep photometric data set over 20 contiguous, equal-width ($\sim$300\AA), non-overlapping, medium-band optical filters (3500\AA -- 9700\AA) plus three standard broad-band near-infrared (NIR) filters ($J$, $H$, and $K_{\rm s}$) over eight different regions of the northern sky \citep{alhambra}. The survey has the aim of understanding the evolution of galaxies throughout cosmic time by sampling a large cosmological fraction of the universe, for which reliable spectral energy distributions (SEDs) and precise photometric redshifts ($z_{\rm p}$) are needed. The final survey parameters and scientific goals, as well as the technical requirements of the filter set, were described by \citet{alhambra}. The survey has collected its data for the 20+3 optical-NIR filters in the 3.5m telescope at the Calar Alto observatory, using the wide-field camera LAICA (Large Area Imager for Calar Alto) in the optical and the OMEGA–2000 camera in the NIR. The full characterisation, description, and performance of the ALHAMBRA optical photometric system were presented in \citet{aparicio10}. A summary of the optical reduction can be found in Crist\'obal-Hornillos et al. (in prep.), the NIR reduction is reported in \citet{cristobal09}.

The wide-field camera LAICA has four chips, each  with a $15\arcmin \times 15\arcmin$ field of view (0.22 arcsec pixel$^{-1}$). The separation between chips is $13\arcmin$. Thus, each LAICA pointing provides four distinct areas in the sky, one per chip. Six ALHAMBRA regions comprise two LAICA pointings. In these cases, the pointings define two separate strips in the sky. We assumed the four chips in each LAICA pointing to be independent sub-fields \citep{clsj14ffcosvar}. We summarise the properties of the seven fields included in the first ALHAMBRA data release\footnote{\url{http://cloud.iaa.es/alhambra/}} in Table~\ref{alhambra_fields_tab}. Currently, ALHAMBRA comprises 48 sub-fields of $\sim183.5$ arcmin$^2$ each.

The sources in the first ALHAMBRA data release were detected in a synthetic $F814W$ filter image, noted $I$ in the following, defined to resemble the HST/$F814W$ filter \citep{molino13}. The areas of the images affected by bright stars and those with lower exposure times (e.g. the edges of the images) were masked following \citet{arnaltemur14}. The total area covered by the current ALHAMBRA data after masking is 2.38 deg$^{2}$ (Table~\ref{alhambra_fields_tab}). Finally, a statistical star/galaxy separation was encoded in the variable \texttt{Stellar\_Flag} of the ALHAMBRA catalogues, and we kept ALHAMBRA sources with $\texttt{Stellar\_Flag} \leq 0.5$ as galaxies. The final catalogue comprises $\sim450$k sources and is complete ($5\sigma$, $3\arcsec$ aperture) for $I \leq 24.5$ galaxies \citep{molino13}.

\subsection{Bayesian photometric redshifts in ALHAMBRA}
The photometric redshifts used throughout were fully presented and tested in \citet{molino13}, and we summarise their principal characteristics below.

The photometric redshifts of ALHAMBRA were estimated with the \texttt{BPZ2} code, a new version of the Bayesian Photometric Redshift \citep[\texttt{BPZ},][]{benitez00} estimator. This is a SED-fitting method based on a Bayesian inference, where a maximum likelihood is weighted by a prior probability. The \texttt{BPZ2} library of 11 SED templates comprises four ellipticals (E, $T \in [1-4]$), one lenticular (S0, $T = 5$), two spirals (S, $T \in [6-7]$), and four starbursts (SB, $T \in [8-11]$). ALHAMBRA relied on the update version of the \texttt{ColorPro} software \citep{colorpro, molino13} to perform point spread function (PSF) matched aperture-corrected photometry, which provided both total magnitudes and isophotal colours for the galaxies. In addition, a homogeneous photometric zero-point recalibration was performed using either spectroscopic redshifts (when available) or accurate photometric redshifts from emission-line galaxies \citep{molino13}. 

The photometric redshift accuracy, as estimated by comparison with $\sim 7200$ spectroscopic redshifts ($z_{\rm s}$), was encoded in the normalised median absolute deviation (NMAD) of the photometric vs spectroscopic redshift distribution \citep{ilbert06,eazy},
\begin{equation}
\sigma_{\rm NMAD} = 1.48 \times \bigg\langle \frac{|\,\delta_z - \langle \delta_z \rangle\,|}{1 + z_{\rm s}} \bigg\rangle,
\end{equation} 
where $\delta_z = z_{\rm s} - z_{\rm p}$ and $\langle \cdot \rangle$ is the median operator. The fraction of catastrophic outliers $\eta$ was defined as the fraction of galaxies with $|\,\delta_z\,|/(1 + z_{\rm s}) > 0.2$. In the case of ALHAMBRA, $\sigma_{\rm NMAD} = 0.011$ for $I \leq 22.5$ galaxies with a fraction of catastrophic outliers of $\eta = 2.1$\%. We refer to \citet{molino13} for a more detailed discussion of the ALHAMBRA photometric redshifts.

\subsection{Probability distribution functions in ALHAMBRA}\label{pdfs}
This section is devoted to the description of the probability distribution functions of the ALHAMBRA sources, those describing the $I-$band magnitude, the photometric redshift, and the quiescent or star-forming classification. These posterior PDFs were needed to successfully compute the luminosity function.

\subsubsection{$I-$band magnitude PDF}\label{ipos}
The ALHAMBRA catalogue was selected in the $I$ band \citep{molino13} and any ALHAMBRA result is affected by this initial selection, even if an absolute magnitude or stellar mass study is performed. Usually, the observed magnitude of selection is assumed without uncertainties both in photometric and spectroscopic surveys, but it is affected by photometric errors. Indeed, we were not interested in the observed $I-$band magnitude of the ALHAMBRA sources, but in their real magnitude, noted $I_0$, unaffected by photometric errors and incompleteness. 

To deal with the $I-$band selection, we defined the posterior PDF of the real $I_0$ magnitude as
\begin{equation}
{\rm PDF}\,(I_0\,|\,I, \sigma_{I}) \propto C(I_0)\ P\,(I\,|\,I_0,\sigma_{I})\label{iopdf},
\end{equation}
where the posterior probability is normalised to unity, $C(I_0)$ is the galaxy number counts [${\rm deg}^{-2}\,{\rm mag}^{-1}$] in the $I$ band, and $P\,(I\,|\,I_0,\sigma_I)$ the probability of observe $I$ having a real magnitude $I_0$ and a photometric error $\sigma_I$. We detail these terms in the following.

\begin{figure}[t]
\centering
\resizebox{\hsize}{!}{\includegraphics{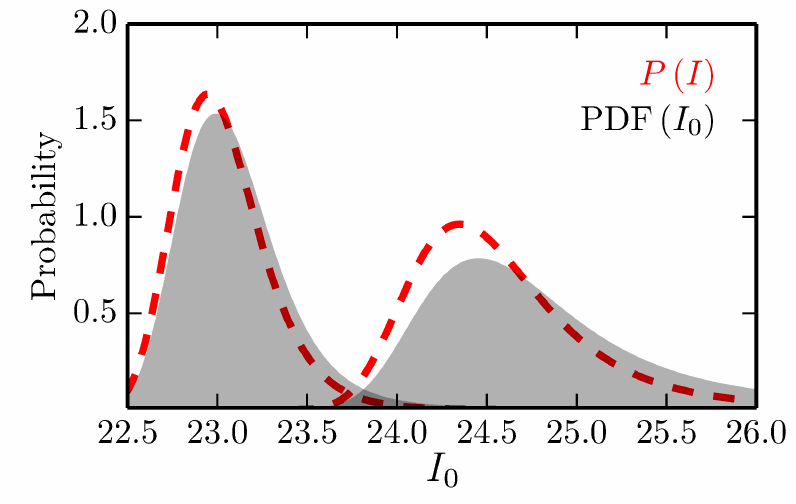}}
\caption{$I$-band magnitude probability $P\,(I)$, red dashed lines, and the posterior ${\rm PDF}\,(I_0)$, grey areas, for two sources with observed magnitude $I = 23$ and $I = 24.5$, and photometric errors $\sigma_I = 0.25$ mag and $\sigma_I = 0.45$ mag, respectively. The number counts prior translates probability to fainter magnitudes.}
\label{pio}
\end{figure}

Photometric errors are Gaussian in flux space and thus asymmetric in magnitude space. Indeed, the probability $P\,(I\,|\,I_0,\sigma_I)$ in magnitude space is
\begin{equation}
P\,(I\,|\,I_0,\sigma_I) = \frac{10^{-0.4(I_0 - I)}}{\sqrt{2\pi}\sigma_I} \exp \bigg\{ -\frac{[1 - 10^{-0.4(I_0 - I)}]^2}{1.7\sigma^2_I} \bigg\}.
\end{equation}
The photometric error $\sigma_I$ was estimated as
\begin{equation}
\sigma_I = \sqrt{\sigma_{\rm phot}^2 + \sigma_{\rm sky}^2 + \sigma_{\rm ZP}^2},
\end{equation}
where $\sigma_{\rm ZP} = 0.02$ is the uncertainty in the zero point, $\sigma_{\rm phot}$ the photon counting error, and $\sigma_{\rm sky}$ the sky background uncertainty. The last was estimated empirically by placing random apertures across the empty areas of the ALHAMBRA images \citep{molino13}. We present two examples of the probability $P\,(I\,|\,I_0,\sigma_I)$ in Fig.~\ref{pio}.

The number counts $C\,(I_0)$ were needed to account for the larger number of faint galaxies and to define a posterior probability \citep[e.g.][]{hogg98, coppin06}. Without this term, we were assuming that galaxies are homogeneously distributed in magnitude space, which is obviously false. The ALHAMBRA $I-$band number counts are presented in Molino et al. (in prep.) and are well described as
\begin{equation}
\log_{10} C\,(I_0) \propto -0.015 I_0^2 + 1.00\,I_0.
\end{equation}
This parametrisation describes well the number counts from $I = 12$ \citep{yasuda01} to $I = 27$ \citep{metcalfe01} and was estimated only with ALHAMBRA data. Following with the example in Fig.~\ref{pio}, the number counts term translates probability to fainter magnitudes.

The posterior ${\rm PDF}\,(I_0)$ was the starting point to define the source function $S$. This function provides the number of sources, corrected by incompleteness and selection effects, with a real magnitude $I_0$ given an observed magnitude $I$ with uncertainty $\sigma_I$. The source function is defined as
\begin{equation}
S(I_0\,|\,I, \sigma_{I}) = \frac{1}{f_{\rm c}\,(I_0)}\,{\rm PDF}\,(I_0|\,I, \sigma_{I}) \int \!\! P\,(I\,|\,I_0,\sigma_{I})\ {\rm d}I_0,\label{sourcefunc}
\end{equation}
where $f_{\rm c}$ is the completeness function and the integral term provides the probability that the source has a positive flux. The last term is smaller than unity only for large uncertainties in the photometry. For example, $\sigma_{I} = 0.5$ mag implies a positive flux probability of 0.98, and $\sigma_{I} = 1$ mag a probability of 0.86. 

\begin{figure}[t]
\centering
\resizebox{\hsize}{!}{\includegraphics{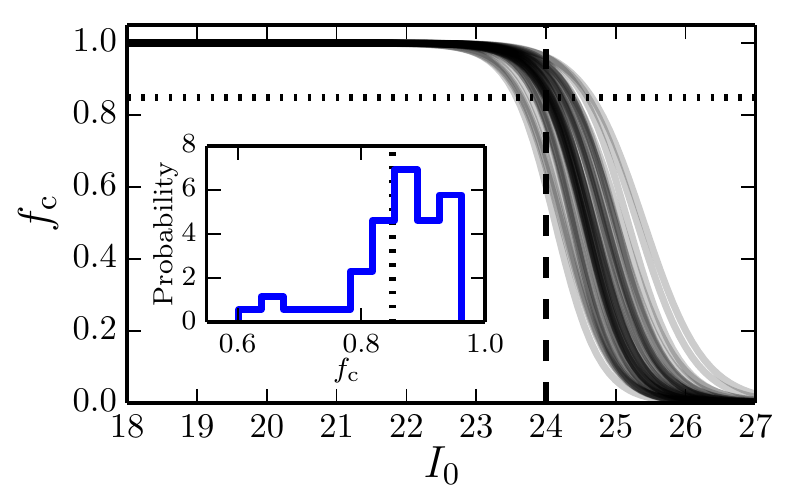}}
\caption{Completeness functions of the 48 ALHAMBRA sub-fields (grey lines). The vertical dashed line marks the selection magnitude $I_0 = 24$ used in the present work. {\it Inset panel} : Normed histogram of the 48 ALHAMBRA sub-fields completeness level at $I_0 = 24$. The dotted lines mark the 85\% completeness in both panels.}
\label{alcomp}
\end{figure}

The completeness function $f_{\rm c}\,(I_0)$ was estimated in each ALHAMBRA sub-field by injecting sources of known $I_0$ magnitude in the $I$-band images and computing their detection rate. As explained in Molino et al (in prep.), to make this estimation as realistic as possible, we preferred not to use point sources but real galaxies (with different shapes, sizes, and magnitudes) extracted from the HST/$F814W$ COSMOS field images \citep{capak07}. The detection rate was fitted with a function of the form
\begin{equation}
f_{\rm c}\,(I_0\,|\,I_{\mu}, \kappa) = \frac{1}{1 + \exp [-\kappa\,(I_0 - I_{\mu})]},
\end{equation}
where $I_{\mu}$ is the 50\% completeness magnitude and $\kappa$ controls the decay rate in the detection. We note that the completeness function $f_{\rm c}$ can only be applied to the real magnitudes $I_0$ because observed magnitudes $I$ are affected by photometric errors. The completeness functions of the 48 ALHAMBRA sub-fields are shown in Fig.~\ref{alcomp}, illustrating the diversity of depths in the survey. A completeness of $f_{\rm c} = 0.85$ is reached on average at $I_0 \sim 24$, with 68\% of the sub-fields in the range $f_{\rm c} \in (0.78, 0.93)$. We set $I_0 = 24$ as our selection magnitude in the following.

We stress here the implications of our real magnitude $I_0 = 24$ selection. This selection was performed {\it a posteriori} in our analysis, in contrast with the {\it a priori} selection in observed $I$ magnitude usually applied in the literature. This is, we included all the ALHAMBRA galaxies in our analysis, even those with $I > 24$, and weighted each with its probability of have a real magnitude $I_0 \leq 24$. This provides 100\% complete samples and a controlled selection function. Thus, with the source function $S(I_0)$ defined in this section, we robustly deal with the $I-$band ALHAMBRA selection, ensuring an unbiased and complete analysis of galaxies selected by their real magnitude $I_0$.

\subsubsection{Photometric redshift PDF}\label{zpos}
\begin{figure}[t]
\centering
\resizebox{\hsize}{!}{\includegraphics{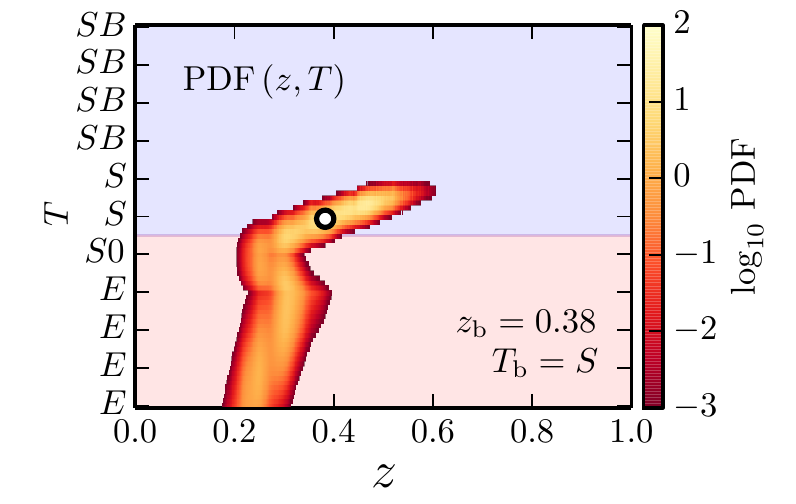}}
\resizebox{\hsize}{!}{\includegraphics{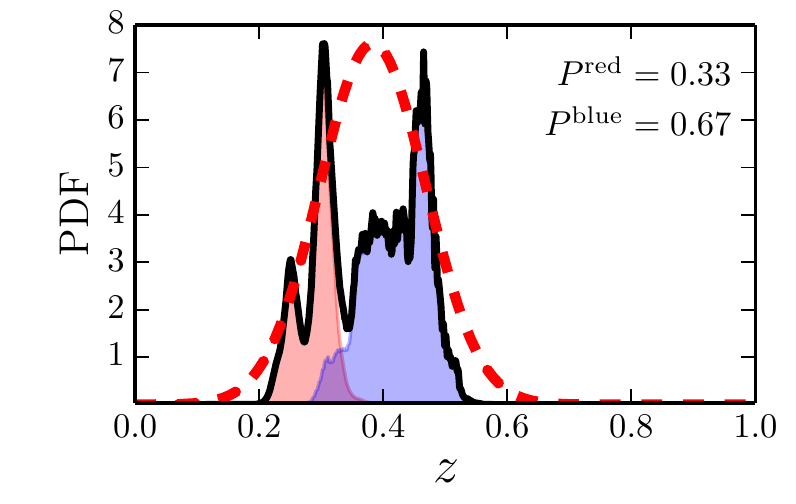}}
\caption{{\it Top panel}: Probability distribution function in the redshift - spectral template ($z - T$) space of an ALHAMBRA galaxy with observed magnitude $I = 22.17 \pm 0.06$. The white dot marks the best Bayesian redshift and template, labelled in the panel. The red area marks `red' spectral templates (${\rm E/S0}$), and the blue area the `blue' spectral templates (${\rm S/SB}$). {\it Bottom panel}: Projection of the top ${\rm PDF}\,(z,T)$ in redshift space. The black solid line marks the total ${\rm PDF}\,(z)$, while the red and blue areas mark the contribution of early and late templates, respectively. This galaxy counts 0.33 as red and 0.67 as blue in our statistical analysis. The red dashed line illustrates the poor Gaussian approximation to this PDF.}
\label{pdfzt}
\end{figure}

As already emphasised by several authors (see Sect.~\ref{intro}), photometric redshifts should not be treated as exact estimates, but as PDFs in a bidimensional (redshift vs spectral type) space. Although the PDF of high signal-to-noise detections can be well-approximated by a Gaussian distribution, for faint detections the photometric uncertainties make these distributions highly non-Gaussian and completely asymmetric, enabling multiple solutions to fit the input photometric data equally well (Fig.~\ref{pdfzt}). This problem, known as the colour--redshift degeneracy, makes PDFs the only robust way to track the uncertainties in the observed photometry to the physical properties of interest. In this context, the ALHAMBRA photometric redshift PDFs have been successfully used to study high-redshift ($z > 2$) galaxies \citep{viironen15}, to detect galaxy groups and clusters \citep{ascaso15}, to estimate the merger fraction \citep{clsj15ffpdf}, or to improve the estimation of stellar population parameters \citep{diazgarcia15}. 

The probability that a galaxy $i$ is located at redshift $z$ and has a spectral type $T$ is $\mathrm{PDF}_{i}\,(z,T)$, see top panel in Fig.~\ref{pdfzt}. This probability density function is the posterior provided by \texttt{BPZ2}. The probability that the galaxy $i$ is located at redshift $z$ is then 
\begin{equation}
\mathrm{PDF}_{i}\,(z) = \int \mathrm{PDF}_{i}\,(z,T)\,{\rm d}T.
\end{equation}
The probability density function $\mathrm{PDF}_i\,(z,T)$ is normalised to one by definition, that is, the probability of any galaxy $i$ being found in the whole parameter space is one. Formally, 
\begin{equation}
1 = \int \mathrm{PDF}_{i}\,(z)\,{\rm d}z  = \int\!\!\!\int \mathrm{PDF}_{i}\,(z,T)\,{\rm d}T\,{\rm d}z.
\end{equation}

The methodology developed in the present paper is only valid if the redshift PDFs were properly computed and calibrated. To test the reliability of the redshift PDFs, several authors use the variable
\begin{equation}
\Delta_z = \frac{\delta_z}{\sigma_z} = \frac{2\,(z_{\rm s} - z_{\rm p})}{\sigma_z^{+} - \sigma_z^{-}},
\end{equation}
where $\sigma_z^{-}$ and $\sigma_z^{+}$ define the redshift range centred in $z_{\rm p}$ that enclose 68\% of the PDF \citep{oyaizu08,oyaizu08sdss,cunha09,ilbert09,reis12,tpz}. The variable $\Delta_z$ should be normally distributed with a zero mean and unit variance if the PDFs are a good descriptor for the accuracy of the photometric redshifts. This is the case for the ALHAMBRA PDFs, as shown by \citet{molino13} and \citet{clsj14ffcosvar}. Thus, the redshift PDFs provided by \texttt{BPZ2} are reliable and can be used to compute the ALHAMBRA luminosity function.

\subsubsection{Quiescent and star-forming PDF}\label{colpos}
Our final goal is to study the luminosity function of the quiescent (galaxies without relevant recent star formation episodes) and the star-forming populations. We defined these galaxy populations in two steps. First, we used the spectral template information encoded in the photometric redshift PDFs to statistically define red and blue galaxies. Then, the proper prior probability was applied to account for the dusty star-forming galaxies that contaminate the red sample as derived from the \texttt{BPZ2} templates.

The definition of blue and red galaxies is not a trivial task, and different authors apply different selections that impact their final results and conclusions. This issue is excellently revised and discussed by \citet{taylor15}. They stress that the two galaxy populations present in the local Universe, that the community labels as red and blue, overlap in colour space and strict colour selections are disfavoured. \citet{taylor15} apply a deconvolution method to recover objectively the two different populations in the colour - stellar mass diagram, providing the statistical weight for belonging to each population given the position in such diagram. Following this framework, our definition of red and blue galaxies takes advantage of the profuse information encoded in \texttt{BPZ2} PDFs. Instead of selecting galaxies according to their observed colour or their best spectral template, we split each PDF into red spectral templates ($T = {\rm E/S0}$) and blue spectral templates ($T = {\rm S/SB}$), as illustrated in the bottom panel of Fig.~\ref{pdfzt}. Formally,
\begin{eqnarray}
\mathrm{PDF}_{i}\,(z) &=& \!\! \mathrm{PDF}_{i}\,(z\,|\,{\rm E/S0}) + \mathrm{PDF}_{i}\,(z\,|\,{\rm S/SB})\nonumber\\ &=& \!\!\int_{T \in\,{\rm E/S0}} \!\!\!\!\!\!\!\!\!\!\mathrm{PDF}_{i}\,(z,T)\,{\rm d}T + \int_{T \in\,{\rm S/SB}} \!\!\!\!\!\!\!\!\!\!\mathrm{PDF}_{i}\,(z,T)\,{\rm d}T.
\end{eqnarray}
The total probability that the galaxy $i$ is either red or blue can be estimated as
\begin{align}
\lefteqn{P_i^{\rm red} = \int \mathrm{PDF}_{i}\,(z\,|\,{\rm E/S0})\,{\rm d}z,}\\
\lefteqn{P_i^{\rm blue} = 1 - P_i^{\rm red} = \int \mathrm{PDF}_{i}\,(z\,|\,{\rm S/SB})\,{\rm d}z.}
\end{align}
In practice, the red templates have $T \in [1,5.5]$ and the blue templates have $T \in (5.5,11]$ in the ALHAMBRA catalogues. This probabilistic description of the two galaxy populations under study, that has been successfully applied in recent work \citep{clsj15ffpdf,clsj15bcosvar,infante15}, is a natural consequence of our PDF analysis. We note that the galaxy presented in Fig.~\ref{pdfzt} has an unique set of observed colours that are compatible within errors with a red (E/S0) and a blue (S/SB) solution simultaneously.

The previous statistical red or blue classification accounts for the uncertainties in the observed photometry, but has an important limitation. The template set of \texttt{BPZ2} was constructed to properly cover the colour space of galaxies, but not their physical properties (e.g. age, metallicity, extinction, star formation rate). Because of this, dust reddened star-forming galaxies could be described by the E/S0 templates of \texttt{BPZ2}, and the red population would comprise therefore quiescent and dusty star-forming galaxies. We resolved this limitation thanks to the MUlti-Filter FITing code \texttt{MUFFIT} \citep{diazgarcia15}. The \texttt{MUFFIT} code is specifically performed and optimised to deal with multi-photometric data, such as the ALHAMBRA dataset, through the SED-fitting (based in a $\chi^2$-test weighted by errors) to mixtures of two single stellar populations (a dominant old component plus a posterior star formation episode, which can be related with a burst or a younger/extended tail in the star formation history). The \texttt{MUFFIT} code includes an iterative process for removing those bands that may be affected by strong emission lines, being able to carry out a detailed analysis of the galaxy SED even when strong nebular or AGN emission lines are present. From \texttt{MUFFIT} analysis, \citet{diazgarcia15} retrieved ages, metallicities, stellar masses, rest-frame luminosities, and extinctions of ALHAMBRA sources with $I \leq 23$. These retrieved parameters are in good agreement with both spectroscopic diagnostics from SDSS data and photometric studies in the COSMOS survey with shared galaxy samples \citep{diazgarcia15}. 

\begin{figure}[t]
\centering
\resizebox{\hsize}{!}{\includegraphics{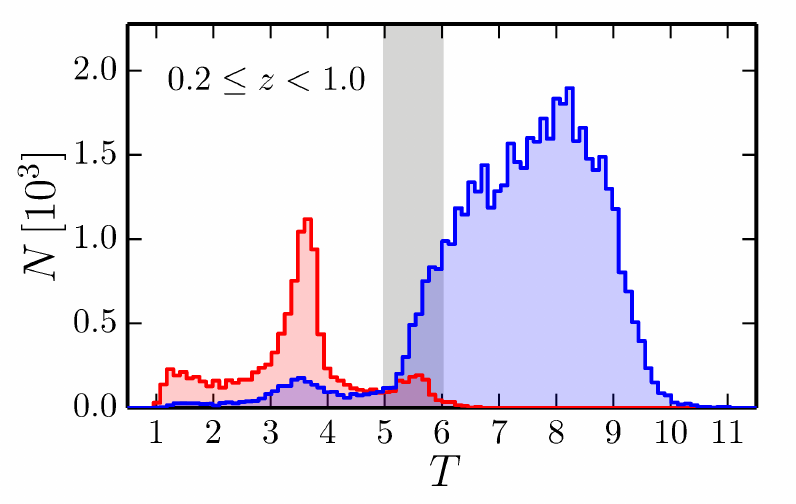}}
\caption{Histogram of the best \texttt{BPZ2} template for the ALHAMBRA $I \leq 23$ galaxies at $0.2 \leq z < 1.0$ classified as quiescent (red) and star-forming (blue) by \texttt{MUFFIT}. The grey area marks the transition region between the E/S0 and S/SB templates of \texttt{BPZ2}.}
\label{histt}
\end{figure}

The position of galaxies in the $UVJ$ colour-colour diagram can be used to select quiescent and star-forming galaxies \citep{williams09, moresco13}. We constructed the ALHAMBRA dust de-reddened $UVJ$ colour-colour diagram with the rest-frame luminosities and the extinction values from \texttt{MUFFIT}, finding that quiescent and star-forming galaxies populates two non-overlapping regions when the effect of dust is accounted for. To test the performance of the \texttt{BPZ2} templates, we used the quiescent or star-forming classification from \texttt{MUFFIT} (D\'{\i}az-Garc\'{\i}a et al., in prep.) We show the distribution of the best \texttt{BPZ2} spectral template $T_{\rm b}$ for the \texttt{MUFFIT} quiescent and star-forming populations in Fig.~\ref{histt}. We find that (i) quiescent galaxies have mainly assigned to E/S0 templates and star-forming galaxies to S/SB templates, as desired. (ii) The transition zone between red and blue templates, $T \in (5,6)$, is populated by quiescent and star-forming galaxies, as expected because of colours uncertainties, but no quiescent galaxy was assigned to S/SB templates. And (iii) there are star-forming galaxies assigned to E/S0 templates, confirming the presence of dusty galaxies in the red population. We studied and parametrised the contamination due to dusty galaxies, defining the probability of being a quiescent (Q) or a dusty star-forming (SF) red galaxy as
\begin{align}
\lefteqn{P\,({\rm Q}\,|\,{\rm E/S0}) = -0.097\,(I_0 - 21) + 0.242\,(z - 0.5) + 0.863,}\\
\lefteqn{P\,({\rm SF}\,|\,{\rm E/S0}) = 1 - P\,({\rm Q}\,|\,{\rm E/S0}),}
\end{align}
where both probabilities are at most unity and at least zero. These probabilities were used as priors in the estimation of the quiescent and star-forming luminosity functions, and are similar to the statistical weights defined by \citet{taylor15} in the colour - stellar mass diagram. We discuss their impact in our results in Sect.~\ref{priorparam}.

Thanks to the probability functions defined in the last sections, we were able to statistically use the output of current photometric redshift codes without losing information and to reliably work with any pre-selection of the sources, neither in the $I$-band magnitude nor in colour.

\section{Estimation of the luminosity function by PDF analysis}\label{lfal}
In this section we detail the steps to compute the posterior luminosity function in ALHAMBRA using the redshift -- spectral template (Sects.~\ref{zpos} and \ref{colpos}) and the $I$-band magnitude posteriors (Sect.~\ref{ipos}). We first derive the $z - M_B$ posterior of each ALHAMBRA source in Sect.~\ref{zmbpos}, and combine them in Sect.~\ref{lfpos} to estimate the ALHAMBRA luminosity function. The procedure to estimate the covariance matrix of the luminosity function, including shot noise and cosmic variance uncertainties, is explained in Sect.~\ref{covalf}. We present the estimation of the galaxy bias function and its covariance matrix in Sects.~\ref{biasfunc} and \ref{covabias}, respectively. Finally, the modelling process followed to describe the observed luminosity and galaxy bias function is detailed in Sect.~\ref{fitting}.

\begin{figure}[t]
\centering
\resizebox{\hsize}{!}{\includegraphics{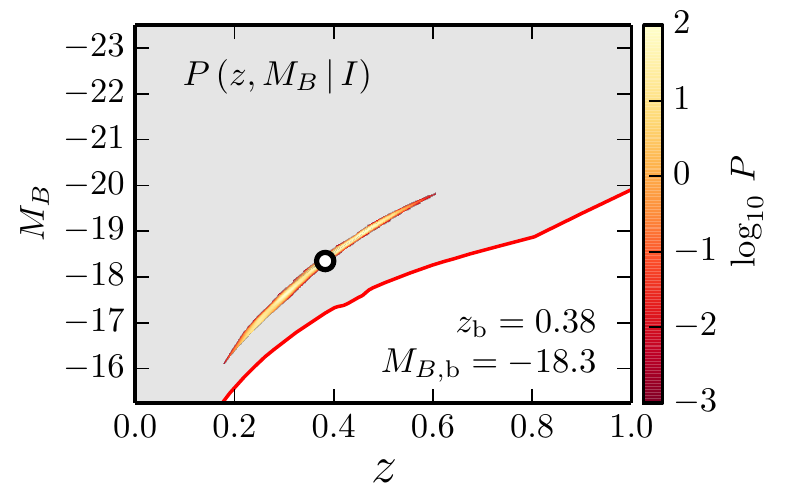}}
\resizebox{\hsize}{!}{\includegraphics{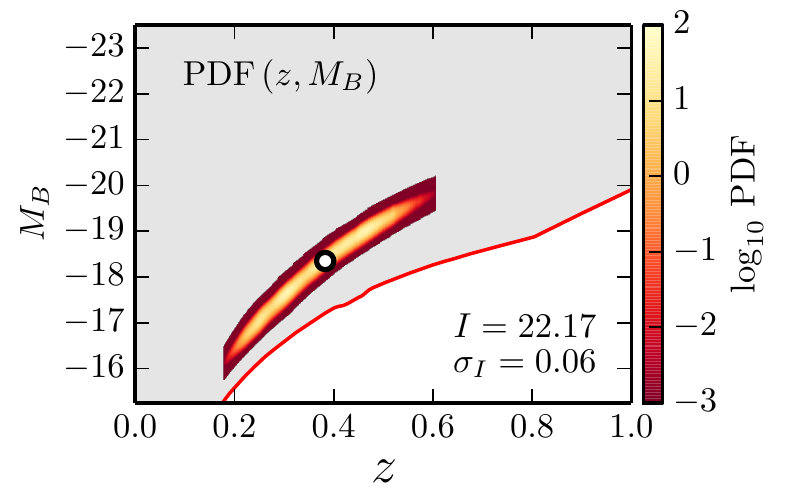}}
\caption{{\it Top panel}: Probability in the $z - M_B$ space, $P\,(z,M_B\,|\,I)$, of the ALHAMBRA galaxy presented in Fig.~\ref{pdfzt}. The white dot marks the best Bayesian redshift and $M_B$, labelled in the panel. {\it Bottom panel}: Posterior probability in the $z - M_B$ space, ${\rm PDF}\,(z,M_B)$, of the same ALHAMBRA galaxy. The convolution with the source function $S(I_0\,|\,I,\sigma_I)$ produces the desired posterior in real magnitude $I_0$. The red solid line in both panels shows the $I_0 = 24$ limiting magnitude, $M_{B,{\rm lim}}$, and the grey areas mark the accessible volumes in $z - M_B$ space.}
\label{pdfzmb}
\end{figure}

\subsection{$z-M_B$ posterior}\label{zmbpos}
The first step in the estimation of the luminosity function is to translate the posterior in the $z - T$ space to the posterior in the $z - M_B$ space. We note that, for a fixed $z$ and $T$, the luminosity distance and the $k$-correction are always the same. Thus, we can map the relation between redshift and spectral template with the $B-$band absolute magnitude $M_B$ using the function $M_B\,(z,T)$, defined as
\begin{equation}
M_B\,(z,T\,|\,I_0) = I_0 - 5\log_{10}[{\rm D_L}(z)] - k\,(z,T) - 25,\label{mbzteq}
\end{equation}
where ${\rm D_L}(z)$ is the luminosity distance in Mpc and $k\,(z,T)$ accounts for the $k$-correction between the observed $I$ band at redshift $z$ and the targeted $B$ band at rest-frame. The estimation of the $k$-correction is detailed in Appendix~\ref{kcorr}. We constructed the probability $P_i\,(z,M_B\,|\,I)$ of each ALHAMBRA source as the ${\rm PDF}$ weighted histogram of $M_{B,i} = M_B\,(z,T\,|\,I_i)$,
\begin{equation}
P_i\,(z,M_B\,|\,I_i)\,{\rm d}M_{B} = \int \mathbf{1}_{M_B}(M_{B,i})\,{\rm PDF}_i\,(z,T)\,{\rm d}T,
\end{equation}
where $\mathbf{1}_{M_B}$ is the indicator function with value unity if the argument is between $M_B$ and $M_B + {\rm d}M_B$. This probability tracks the uncertainties of the observed colours to the $z - M_B$ space, including the correlation between both variables. We present the $P\,(z,M_B\,|\,I)$ of the Fig.~\ref{pdfzt} ALHAMBRA source in the upper panel of Fig.~\ref{pdfzmb}. Nevertheless, this probability is not the desired $z - M_B$ posterior because it was estimated using the observed magnitude $I$. We computed the final posterior ${\rm PDF}\,(z,M_B)$ by convolving the previous probability with the source function defined in Sect.~\ref{ipos},
\begin{equation}
{\rm PDF}_i\,(z,M_B) = P_i\,(z,M_B\,|\,I_i) \ast S(I_0\,|\,I_i, \sigma_{I,i}).
\end{equation}
This procedure includes in the final posterior the uncertainties in the flux normalization of the source, as shown in the lower panel of Fig.~\ref{pdfzmb}.

Our final goal is the study of the quiescent and star-forming luminosity function. With the template information encoded in the photometric redshift PDFs and the quiescent or star-forming probability for red galaxies derived in Sect.~\ref{colpos}, we computed the desired posteriors of quiescent (Q) and star-forming (SF) galaxies as
\begin{align}
{\rm PDF}_i\,&(z,M_B\,|\,{\rm Q}) = \nonumber\\
& P_i\,(z,M_B\,|\,I_i, {\rm E/S0}) \ast [S(I_0\,|\,I_i, \sigma_{I,i}) \times P\,({\rm Q\,|\,E/S0})]
\end{align}
and
\begin{align}
{\rm PDF}_i\,&(z,M_B\,|\,{\rm SF}) = \nonumber\\
& P_i\,(z,M_B\,|\,I_i, {\rm E/S0}) \ast [S(I_0\,|\,I_i, \sigma_{I,i}) \times P\,({\rm SF\,|\,E/S0})] + \nonumber\\
& P_i\,(z,M_B\,|\,I_i, {\rm S/SB}) \ast S(I_0\,|\,I_i, \sigma_{I,i}).
\end{align}
In the previous equations the quiescent or star-forming probability is a function of $I_0$ and $z$, and it was applied to the source function $S$ at each $z$ before the convolution.

\begin{figure}[t]
\centering
\resizebox{\hsize}{!}{\includegraphics{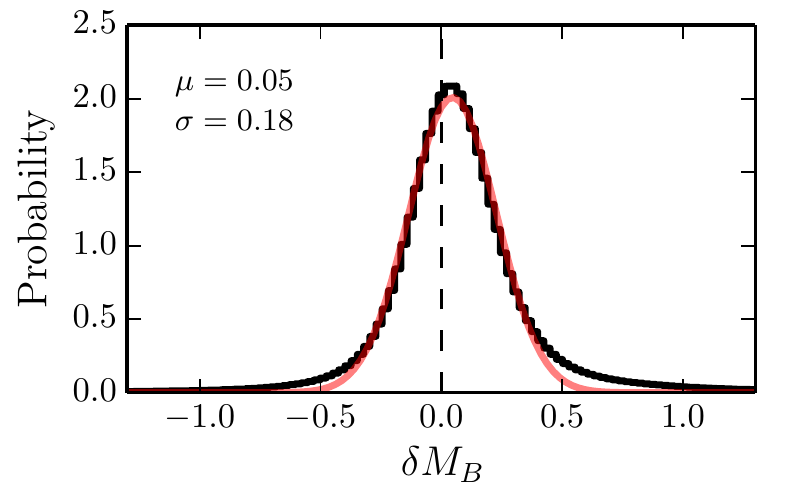}}
\caption{Difference between the $M_B$ posterior derived from \texttt{BPZ2} and the $B-$absolute magnitude provided by \texttt{MUFFIT} for a common sample of $I \leq 23$ ALHAMBRA galaxies at $0.2 \leq z < 1.0$ (black histogram). The red solid curve shows the best Gaussian fit to the distribution, with median $\mu = 0.05$ mag and dispersion $\sigma = 0.18$ mag. The dashed line marks identity.}
\label{mbmuffit}
\end{figure}

To ensure the reliability of the \texttt{BPZ2} absolute magnitudes computed in this section, we compared the derived $M_B$ posterior, defined as
\begin{equation}
{\rm PDF}\,(M_B) = \int {\rm PDF}\,(z,M_B)\,{\rm d}z,
\end{equation}
with the $B-$band absolute magnitude estimated by \texttt{MUFFIT}, noted $M_{B}^{\texttt{MUFFIT}}$. We show the comparison between \texttt{BPZ2} and \texttt{MUFFIT} at $0.2 \leq z < 1.0$ in Fig.~\ref{mbmuffit}, estimated with the variable
\begin{equation}
\delta M_B = \sum_i {\rm PDF}_i\,(M_B) - M_{B,i}^{\texttt{MUFFIT}}.
\end{equation}
We find that $\delta M_B$ follows a Gaussian distribution with mean $\mu = 0.05$ mag and dispersion $\sigma = 0.18$ mag. We explored different redshift ranges both for quiescent and star-forming galaxies, and we find that the differences between \texttt{BPZ2} and \texttt{MUFFIT} are $< 0.1$ mag in any case, with a typical dispersion of $\sigma \sim 0.18$ mag. Because both codes were applied over the same photometric dataset, the expected uncertainty of each code individually is $\sigma/\!\sqrt{2} \sim 0.13$ mag. From the width of the derived ${\rm PDF}\,(M_B)$, we estimated $\sigma_{\texttt{BPZ}} \sim 0.12$ mag, and from the \texttt{MUFFIT} results we find $\sigma_{\texttt {MUFFIT}} \sim 0.12$ mag. Both uncertainties are similar and close to the expected one. Because of the small offset with respect to \texttt{MUFFIT} and the well behaved uncertainties, we conclude that the \texttt{BPZ2} $B$-band absolute magnitudes and their errors are reliable, and we can use therefore the posterior ${\rm PDF}\,(z,M_B)$ to compute the luminosity function.

\subsection{Luminosity function by PDF analysis}\label{lfpos}
As demonstrated by \citet{sheth10}, the real luminosity function in photometric surveys can be constructed with the posterior ${\rm PDF}\,(z,M_B)$ estimated in the previous section. The posterior luminosity function of the ALHAMBRA sub-field $j$ was measured as
\begin{equation}
\Phi_j\,(z,M_B) = \frac{1}{A_j} \sum_i \mathrm{PDF}_{i}\,(z,M_B)\,\bigg( \frac{{\rm d}V^{\prime}}{{\rm d}z} \bigg)^{-1}\,\,\,\,[{\rm Mpc}^{-3}{\rm mag}^{-1}],
\end{equation}
where the index $i$ runs the galaxies in the sub-field, ${\rm PDF}_{i}\,(z,M_B)$ is the posterior in the redshift -- absolute magnitude space of galaxy $i$, $A_j$ the area subtended by the sub-field $j$ in deg$^{2}$, and ${\rm d}V^{\prime} / {\rm d}z$ the differential cosmic volume probed by one square degree, defined as
\begin{equation}
\frac{{\rm d}V^{\prime}}{{\rm d}z} = \frac{\pi^2}{180^2} \frac{c}{H_0} \frac{(1+z)^2 D_A^2(z)}{E(z)}\,\,\,\,\,[{\rm Mpc}^{3}\,{\rm deg}^{-2}],
\end{equation}
where $c$ is the speed of light, $D_A(z)$ the angular diameter distance, and $E(z)~=~\sqrt{\Omega_{\rm m}(1+z)^3 + \Omega_{\Lambda}}$. 

We are interested on the study of star-forming and quiescent galaxies, so we computed
\begin{eqnarray}
\lefteqn{\Phi_j^{\rm SF}\,(z,M_B) = \Phi_j\,(z,M_B\,|\,{\rm SF}),}\\ 
\lefteqn{\Phi_j^{\rm Q}\,(z,M_B) = \Phi_j\,(z,M_B\,|\,{\rm Q}).}
\end{eqnarray}
We note that
\begin{equation}
\Phi^{\rm tot}_j\,(z,M_B) = \Phi_j^{\rm SF}\,(z,M_B) + \Phi_j^{\rm Q}\,(z,M_B) = \sum_t \Phi_j^t\,(z,M_B),
\end{equation}
where the index $t$ runs the two galaxy populations under study. These luminosity functions were computed for galaxies with real $I-$band magnitude brighter than $I_0 = 24$ (see Sect.~\ref{ipos}, for details). We estimated the limiting $M_B$ at each redshift as the $B-$band absolute magnitude of the brighter template $T$,
\begin{equation}
M_{B,{\rm lim}}(z) = {\rm min}\,[\,M_B\,(z,T\,|\,I_0 = 24)\,].
\end{equation}
Because we were working on real magnitudes thanks to the $I$-band source function, the limiting $M_B$ translates to 100\% completeness both for star-forming and quiescent galaxies in all the explored ranges of luminosity and redshift. We show $M_{B,{\rm lim}}$ in both panels of Fig.~\ref{pdfzmb}.

To ensure a well controlled error budget of the luminosity function (see Sect.~\ref{binning}, for details), we degraded the resolution of each $\Phi^t_j\,(z,M_B)$ to create the binned luminosity function,
\begin{align}
\tens{\Phi}^t_j \equiv \Phi^t_j\,&(\vec{z},\vec{M_B}) \equiv \Phi^t_j\,(z_m,M_{B,n}) = \nonumber\\
& \frac{1}{\Delta V_m\,\Delta M_{B,n}} \int_{z^{-}_m}^{z^{+}_m} \!\!\! \int_{M_{B,n}^{-}}^{M_{B,n}^{+}} \!\! \Phi^t_j\,(z,M_B) \frac{{\rm d}V^{\prime}}{{\rm d}z}\,{\rm d}z\,{\rm d}M_B,\label{binlf}
\end{align}
where $\mathbf{z}$ and $\mathbf{M_{B}}$ are the vectors that define the binned histogram, $z^{-}_m = z_m - 0.5\Delta z_m$, $z^{+}_m = z_m + 0.5\Delta z_m$, $M_{B,n}^{-} = M_{B,n} - 0.5\Delta M_{B_n}$, and $M_{B,n}^{+} = M_{B,n} + 0.5\Delta M_{B,n}$ the integration limits of the bins, $\Delta z_m$ and $\Delta M_{B,n}$ the bin size vectors in redshift and $B-$band absolute magnitude, respectively, and $\Delta V_m$ the cosmic volume probed by one square degree at $z^{-}_m \leq z < z^{+}_m$,
\begin{equation}
\Delta V_{m} = \int_{z_m^-}^{z_m^+} \frac{{\rm d}V^{\prime}}{{\rm d}z}\,{\rm d}z.
\end{equation}
We define the optimum bin sizes and justify the need of binned luminosity functions in Sect.~\ref{binning}.

We combined the ALHAMBRA sub-fields to obtain the total binned ALHAMBRA $B-$band luminosity function,
\begin{equation}
\tens{\Phi}^{\rm tot} = \tens{\Phi}^{\rm SF} + \tens{\Phi}^{\rm Q} = \sum_t \tens{\Phi}^t = \sum_t \frac{1}{N} \sum_j \tens{\Phi}^t_j,
\end{equation}
where the index $j$ runs the $N = 48$ ALHAMBRA sub-fields. We defined the ALHAMBRA luminosity function with the tensor $\tens{\Phi} = [\tens{\Phi}^{\rm SF}, \tens{\Phi}^{\rm Q}]$. These equations are also valid to obtain the final differential ALHAMBRA luminosity function. We present the uncertainty estimation of $\tens{\Phi}$ in Sect.~\ref{covalf}.

\subsection{Galaxy bias function}\label{biasfunc}
In the previous section, we described the estimation of the galaxy distribution average in $z - M_B$ space, the luminosity function. Taking advantage to the several sub-fields of the ALHAMBRA survey, we had also access to the dispersion of such distribution. As shown by \citet{clsj15bcosvar}, the galaxy bias $b_v$ can be estimated from the intrinsic dispersion of the galaxy distribution (i.e. the cosmic variance $\sigma_v$) by comparison with the cosmic variance of the dark matter predicted by the theory. The galaxy bias is the relationship between the spatial distribution of galaxies and the underlying dark-matter density field \citep{kaiser84, bardeen86, mo96}.

The galaxy bias measured with the cosmic variance in ALHAMBRA agrees with the bias estimated by correlation function analysis from \citet{arnaltemur14} over the same data set, and in addition to the luminosity function we also estimated the ALHAMBRA galaxy bias function $b_{v}\,(z,M_B)$. This bias function was used to estimate the cosmic variance term of the luminosity function covariance matrix (Sect.~\ref{cosvar}) and provides hints about the interplay between galaxies and dark matter.

We noticed that the galaxy bias measurements presented on \citet{clsj15bcosvar} are computed for the total population in overlapping redshift ranges on samples selected with a luminosity threshold. To estimate the bias function of quiescent and star-forming galaxies on independent bins, we defined several non-overlapping volumes in the $z - M_B$ space. Then, we measured the cosmic variance $\sigma_v$ and its associated galaxy bias following the methodology described in \citet{clsj15bcosvar}. The galaxy bias function of the population $t$ is defined as
\begin{equation}
\tens{b}^t_v \equiv b^t_v\,(\mathbf{z},\mathbf{M_B}) = \frac{\sigma^t_v\,(\mathbf{z},\mathbf{M_B})}{\sigma_{v,{\rm dm}}\,(\mathbf{z})},
\end{equation}
where $\sigma_{v,{\rm dm}}\,(\mathbf{z})$ is the cosmic variance of the dark matter predicted by the theory at $z_m \pm \Delta z_m$ for a subtended area of $\langle A_j \rangle = 0.051$ deg$^2$, the median area of the 48 ALHAMBRA sub-fields. The theoretical cosmic variance was computed in each volume using the code \texttt{QUICKCV}\footnote{\texttt{QUICKCV} is available at \url{www.phyast.pitt.edu/~janewman/quickcv}}, which is described in \citet{quickcv}. The code computes the cosmic variance from the dark-matter power spectrum using a window function which is 1 inside the interest volume and 0 otherwise. We obtained the dark-matter power spectrum at each redshift bin using the \texttt{CAMB} software \citep{camb}, including the non-linear corrections of \texttt{HALOFIT} \citep{halofit}. We define the volumes used in the estimation of the galaxy bias function in Sect.~\ref{binning}.

\subsection{Luminosity function covariance matrix}\label{covalf}
The uncertainty in the luminosity function has two dominant terms \citep{robertson10lf,smith12}: the statistical error (i.e, the shot noise) and the cosmic variance. In this section we describe the estimation of the ALHAMBRA luminosity function covariance matrix, both the shot noise (Sect.~\ref{poisson}) and cosmic variance (Sect.~\ref{cosvar}) terms. 

\subsubsection{Shot noise term}\label{poisson}
Because of the uncertainties in the photometric redshifts and the observed magnitudes, the luminosity function values of adjacent bins are correlated in both dimensions. Moreover, quiescent and star-forming luminosity functions are also correlated because of the $z - T$ degeneracies, as shown in Fig.~\ref{pdfzt}. We estimated the shot noise term of the $\tens{\Phi}$ covariance matrix with the bootstrapping technique \citep{bootstrap}. We created $k = 20000$ bootstrap samples of the luminosity functions in each ALHAMBRA sub-field, noted $\tens{\Phi}^t_{j,k}$, and computed the shot noise term as
\begin{align}
\tens{\Sigma}_{{\rm P},j} \equiv \Sigma_{{\rm P},j}\,(t_1,t_2,&\,z_m,z_p,M_{B,n},M_{B,q}) = \nonumber\\ 
& \frac{\mathbb{E}\,[\Phi^{t_1}_{j,k}\,(z_m,M_{B,n})\,\Phi^{t_2}_{j,k}\,(z_p,M_{B,q})]}{\Phi^{t_1}_j(z_m,M_{B,n})\,\Phi^{t_2}_j(z_p,M_{B,q})} - 1,
\end{align}
where $\mathbb{E}$ is the expected value (i.e. the mean) operator, the indices $t_1$ and $t_2$ run the galaxy types, the indices $m$ and $p$ run the redshift bins, and the indices $n$ and $q$ run the absolute magnitude bins. We map the covariance between luminosity bins at the same redshift and galaxy type when $m = p$ and $t_1 = t_2$, the covariance between redshift bins at the same luminosity and galaxy type when $n = q$ and $t_1 = t_2$, and the covariance between quiescent and star-forming galaxies when $t_1 \neq t_2$. We computed the relative covariance matrix because the number density values are log-normally distributed \citep[e.g.][]{coles91, clsj15bcosvar}, and it is natural to work in log-space.

The shot noise term of the ALHAMBRA luminosity function is
\begin{equation}
\tens{\Sigma}_{\rm P} = \frac{1}{N^2} \sum_j \tens{\Sigma}_{{\rm P},j}.\label{eqsnoise}
\end{equation}
We assumed that the luminosity functions from different sub-fields are independent. The independence approximation in ALHAMBRA is valid for our proposes, as demonstrated by \citet{clsj14ffcosvar}.

\subsubsection{Cosmic variance term}\label{cosvar}
The relative cosmic variance $\sigma_v$ is a fundamental uncertainty in any observational measurement derived from galaxy surveys, arising from the underlying large-scale density fluctuations and leading to variances larger than those expected from the Poisson statistics estimated in the previous section.

To estimate the relative cosmic variance term, we used the galaxy bias functions $\tens{b}^c_v$ defined in Sect.~\ref{biasfunc}. The observational bias function was computed in $z - M_B$ volumes that are larger than those used to compute the luminosity function (see Sect.~\ref{binning}, for details). Because of this, we used the modelled bias function estimated in Sect.~\ref{biasfit} instead of the observed one to derive the cosmic variance term.

We estimated the relative cosmic variance of the galaxy population $c$ as
\begin{equation}
\sigma^t_{v, {\rm mod}}\,(\mathbf{z}, \mathbf{M_B}) = \frac{1}{\sqrt{N}}\,b^t_{v, {\rm mod}}\,(\mathbf{z},\mathbf{M_{B}}) \times \sigma_{v,{\rm dm}}(\mathbf{z}),\label{sigmamod}
\end{equation}
where the factor $\sqrt{N}$ accounts for the combination of the ALHAMBRA sub-fields, and
\begin{align}
\tens{b}^t_{v,{\rm mod}} & \equiv b^t_{v,{\rm mod}}\,(\mathbf{z},\mathbf{M_{B}}) = \frac{1}{\tens{\Phi}^t_{\rm mod}\,\Delta V_m\,\Delta M_{B,n}} \times \nonumber\\
& \int_{z^{-}_m}^{z^{+}_m} \!\! \int_{M_{B,n}^{-}}^{M_{B,n}^{+}} \!\! b^t_{v, {\rm mod}}\,(z,M_B)\,\Phi^t_{\rm mod}\,(z,M_B) \frac{{\rm d}V^{\prime}}{{\rm d}z} {\rm d}z\,{\rm d}M_B,\label{bvbin}
\end{align}
is the binned version of the modelled galaxy bias function $b^t_{v, {\rm mod}}\,(z,M_B)$ described in Sect.~\ref{biasfit}. The modelled luminosity function $\Phi^t_{\rm mod}\,(z,M_B)$ and its binned version $\tens{\Phi}^t_{\rm mod}$ are described in Sect.~\ref{lffit}.

The cosmic variance term of the covariance matrix is then
\begin{align}
\tens{\Sigma}_v \equiv &\,\Sigma_v\,(t_1,t_2,z_m,z_p,M_{B,n},M_{B,q}) = \nonumber\\ 
& \delta_{mp}\,\frac{\sigma^{t_1}_{v, {\rm mod}}\,(z_m, M_{B,n})\,\sigma^{t_2}_{v, {\rm mod}}\,(z_p, M_{B,q})}{\sqrt{V_{\rm eff}\,(z_m, M_{B,n})\,V_{\rm eff}\,(z_p, M_{B,q})}}\,\sqrt{\Delta V_m\,\Delta V_p},
\end{align}
where the Kronecker $\delta_{mp}$ is one if $m = p$ and zero otherwise, and the effective volume $V_{\rm eff}$ is estimated as
\begin{equation}
V_{\rm eff}\,(\mathbf{z},\mathbf{M_{B}}) = \frac{1}{\Delta M_{B,n}} \int_{z_m^-}^{z_m^+} \int^{{\rm min} [M_{B,n}^+,\,M_{B,{\rm lim}}]}_{M_{B,n}^-} \frac{{\rm d}V^{\prime}}{{\rm d}z}\,{\rm d}z\,{\rm d}M_B.
\end{equation}
With the effective volume we account for the lower cosmic volume (i.e. larger cosmic variance) probed in those magnitude bins affected by the $I_0 = 24$ selection. The definition of $\tens{\Sigma}_v$ implies that the redshift bins are independent, and that the luminosities and the galaxy types are highly correlated, that is, an over-dense field has an excess of both quiescent and star-forming galaxies at any luminosity \citep{smith12}.

\subsubsection{Final covariance matrix}
The final relative covariance matrix of the ALHAMBRA luminosity function is
\begin{equation}
\tens{\Sigma}_{\Phi} = \tens{\Sigma}_{\rm P} + \tens{\Sigma}_v. 
\end{equation}
The covariance matrix $\tens{\Sigma}_{\Phi}$ tracks not only the correlations due to the redshift and magnitude uncertainties, but also the correlations due to the cosmic variance that strongly couple the luminosity functions of quiescent and star-forming galaxies. The properties of the estimated covariance matrix are discussed in Sect.~\ref{covdis}. We note that the $N = 48$ ALHAMBRA sub-fields subtend a similar sky area, so the cosmic variance term can be estimated as $1/\sqrt{N}$ of one single sub-field cosmic variance, simplifying the process with Eq.~(\ref{sigmamod}). If sub-fields of different areas are available, the cosmic variance term should be estimated for each individual sub-field and then combined following the prescriptions in \citet{moster11}.

\subsection{Galaxy bias function covariance matrix}\label{covabias}
Following Sect.~\ref{poisson}, we estimated the covariance matrix $\tens{\Sigma}_b^t$ of the galaxy bias function from $k = 100$ bootstrap samples of the $48$ ALHAMBRA sub-fields,
\begin{align}
\Sigma_{b}^{t}\,(z_m,z_p,M_{B,n},M_{B,q}) = &\,\delta_{nq}\,\mathbb{E}\,[b^{t}_{v,k}\,(z_m,M_{B,n})\,b^{t}_{v,k}\,(z_p,M_{B,q})]\nonumber\\ 
& - \delta_{nq}\,b^{t}_v\,(z_m,M_{B,n})\,b^{t}_v\,(z_p,M_{B,q}),
\end{align}
where the Kronecker $\delta_{nq}$ is one if $n = q$ and zero otherwise. Two main differences should be noted with respect to the shot noise term of the luminosity function covariance matrix presented in Sect.~\ref{poisson}. First, the galaxy bias was studied in real space, and thus the relative matrix was not needed. Second, we did not track the covariance between neither quiescent and star-forming galaxies nor different luminosities. This decision was motivated by the origin of the galaxy bias signal, that it is estimated from the intrinsic dispersion (i.e. the cosmic variance) of the ALHAMBRA sub-fields. As argued in Sect.~\ref{cosvar}, the cosmic variance highly correlates quiescent and star-forming galaxies at any luminosity \citep{smith12}, and only independent measurements are expected at different redshifts. Thus, we only should track the observational covariance of $b_v$ between different redshift bins, which is not expected and could therefore impact the fitting process. We estimated that the correlation coefficient in $b_v$ between different $t$ and $M_B$ is in the range $0.4-0.6$, a expected large value that confirms the previous arguments.

\begin{figure}[t]
\centering
\resizebox{\hsize}{!}{\includegraphics{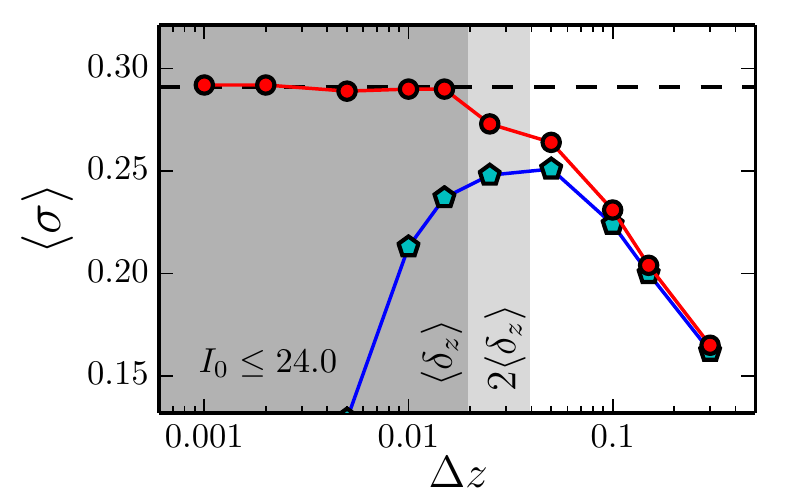}}
\resizebox{\hsize}{!}{\includegraphics{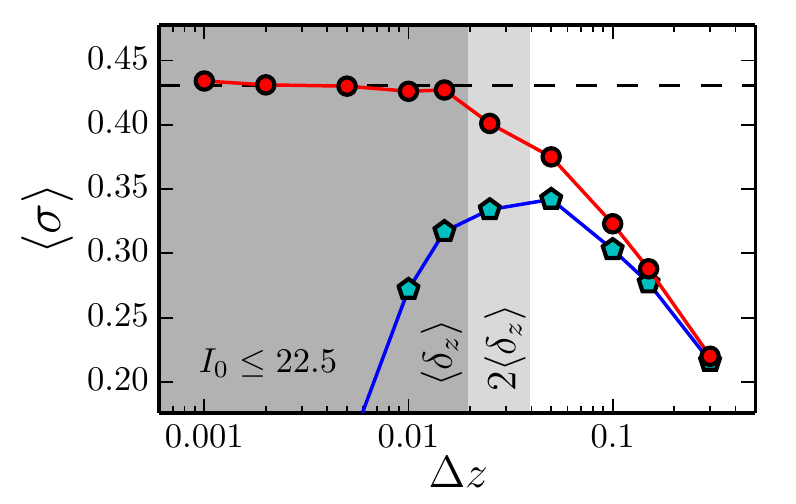}}
\caption{Median total variance (red dots) and the median cosmic variance (cyan pentagons) as a function of the redshift bin size $\Delta z$ for $I_0 \leq 24$ ({\it top panel}) and $I_0 \leq 22.5$ galaxies ({\it bottom panel}) at $0.2 \leq z < 1$. The dashed lines mark the total variance in the constant regime. The grey areas mark those bin sizes smaller than $\langle \delta_z \rangle$, the ALHAMBRA photometric redshift precision, and $2\langle \delta_z \rangle$.}
\label{sig_fig}
\end{figure}

\subsection{Joint modelling of the luminosity and galaxy bias functions}\label{fitting}
The modelling of the ALHAMBRA luminosity and galaxy bias functions is described in the following sections. The presented process is general and we could change the models and their defining parameters in the future. 

\subsubsection{Luminosity function model}\label{lffit}
We modelled the ALHAMBRA luminosity function with the function
\begin{equation}
\Phi_{\rm mod}\,(z, M_{B}\,|\,\boldsymbol{\theta}_{\Phi}) = [\Phi^{\rm SF}_{\rm mod},\Phi_{\rm mod}^{\rm Q}],\label{modeltot}
\end{equation}
where $\boldsymbol{\theta}_{\Phi} = [\boldsymbol{\theta}_{\Phi}^{\rm SF},\boldsymbol{\theta}_{\Phi}^{\rm Q}]$ are the parameters that define the model and which we want to estimate. 

We estimated the posterior distribution of the model parameters $\boldsymbol{\theta}_{\Phi}$ as
\begin{equation}
P\,(\boldsymbol{\theta}_{\Phi}\,|\,\tens{\Phi},\tens{\Sigma}_{\Phi}) \propto {\rm exp}\,\bigg(-\frac{1}{2}\,\chi^2\bigg)\,P\,(\boldsymbol{\theta}_{\Phi}),\label{pbayes}
\end{equation}
where the posterior distribution is normalised to unity, $P\,(\boldsymbol{\theta}_{\Phi})$ is the prior in the parameters, and the $\chi^2$ function is defined as
\begin{equation}
\chi^2\,(\tens{\Phi}\,|\,\boldsymbol{\theta}_{\Phi},\tens{\Sigma}_{\Phi}) = [\ln \tens{\Phi} - \ln \tens{\Phi}_{\rm mod}]^{\rm T}\,\tens{\Sigma}_{\Phi}^{-1}\,[\ln \tens{\Phi} - \ln \tens{\Phi}_{\rm mod}],\label{lfml}
\end{equation}
where $\tens{\Phi}_{\rm mod}$ is the binned version, following Eq.~(\ref{binlf}), of the model $\Phi_{\rm mod}$. The $\chi^2$ function was defined in log-space because the luminosity function values follow a log-normal distribution instead of a Gaussian one. We assumed uninformative priors on the parameters, that is, $P\,(\boldsymbol{\theta}_{\Phi}) = 1$.

We modelled the ALHAMBRA luminosity function with a combination of Schechter functions. The Schechter function is defined with three parameters, the characteristic magnitude $M_{B}^{*}$ (corresponds to the transition magnitude from a power law luminosity function to an exponential one), the characteristic density $\phi^{*}$ (the normalization of the function in Mpc$^{-3}$ mag$^{-1}$, roughly equivalent to the density at $M_{B}^{*}$), and the slope $\alpha$ (determines the slope of the power law variation at the faint end). Formally,
\begin{equation}
\mathcal{S}\,(M_B\,|\,M_{B}^{*},\phi^{*},\alpha) = 0.4\ln(10)\,\phi^{*}\,\frac{10^{0.4[M^{*}_{B} - M_B](1+\alpha)}}{\exp\big\{10^{0.4[M^{*}_{B} - M_B]}\big\}}.
\end{equation}

\begin{figure*}[!t]
\centering
\resizebox{0.49\hsize}{!}{\includegraphics{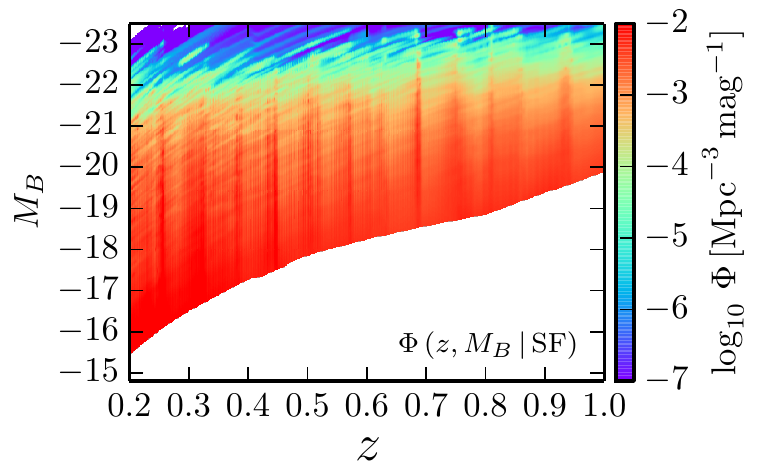}}
\resizebox{0.49\hsize}{!}{\includegraphics{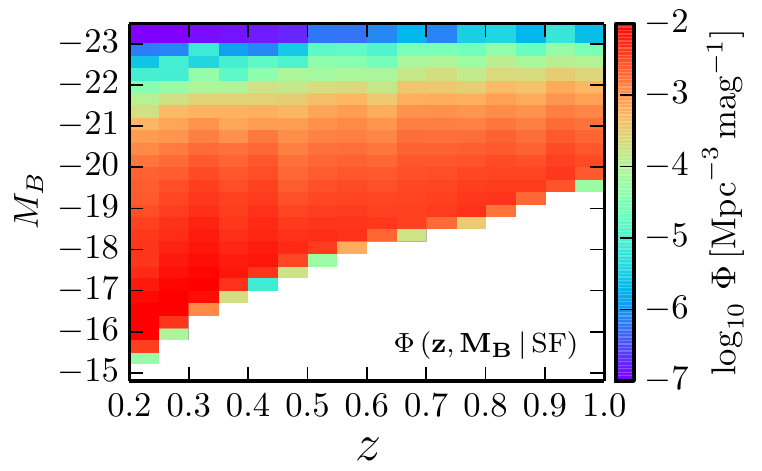}}\\
\resizebox{0.49\hsize}{!}{\includegraphics{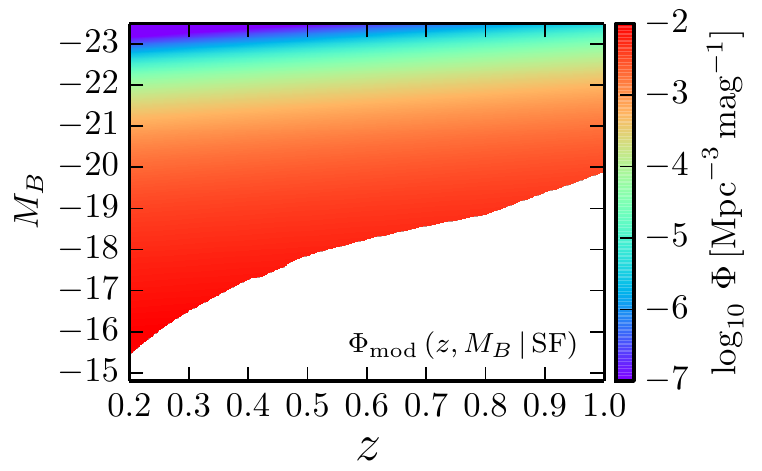}}
\resizebox{0.49\hsize}{!}{\includegraphics{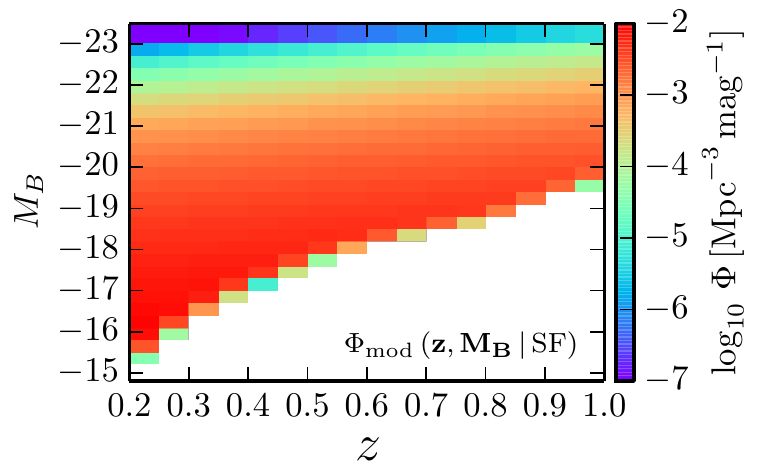}}
\caption{{\it Top panels}: Posterior luminosity function of star-forming galaxies in ALHAMBRA, $\Phi\,(z,M_B\,|\,{\rm SF})$, differential ({\it left}) and binned ({\it right}) version with $\Delta z = 0.05$ and $\Delta M_B = 0.3$. {\it Bottom panels}: Median luminosity function model for star-forming galaxies, differential ({\it left}) and binned ({\it right}) version. These luminosity functions are complete for $I_0 \leq 24$ galaxies.}
\label{lf2d_blue}
\end{figure*}

We defined the model luminosity function for star-forming galaxies as a redshift-dependent Schechter function,
\begin{equation}
\Phi_{\rm mod}^{\rm SF}\,(z,M_B\,|\,\boldsymbol{\theta}^{\rm SF}_{\Phi}) = \mathcal{S}\,(M_B\,|\,M_{B,{\rm SF}}^*(z),\phi^{*}_{\rm SF}(z),\alpha_{\rm SF}),\label{modelblue}
\end{equation}
where
\begin{eqnarray}
\lefteqn{M^{*}_{B,{\rm SF}}\,(z) = M_{B,{\rm SF}}^0 + Q_{\rm SF}\,(z - 0.5),}\label{mstarz}\\
\lefteqn{\log_{10} \phi^{*}_{\rm SF}(z) = \phi^0_{\rm SF} + P_{\rm SF}\,(z - 0.5).}\label{phiz}
\end{eqnarray}
We assumed the faint-end slope $\alpha_{\rm SF}$ as constant with redshift, so $\boldsymbol{\theta}^{\rm SF}_{\Phi}~=~[M_{B,{\rm SF}}^0, Q_{\rm SF}, \phi^0_{\rm SF}, P_{\rm SF}, \alpha_{\rm SF}]$.

We defined the model luminosity function for quiescent galaxies as a combination of two redshift-dependent Schechter functions
\begin{align}
\Phi_{\rm mod}^{\rm Q}\,(z,M_B\,|\,\boldsymbol{\theta}^{\rm Q}_{\Phi}) =&\,\mathcal{S}\,(M_B\,|\,M_{B,{\rm Q}}^*(z),\phi^{*}_{\rm Q}(z),\alpha_{\rm Q}) \nonumber\\
& + \mathcal{S}\,(M_B\,|\,M_{\rm f},\phi^{*}_{\rm Q}(z),\beta),\label{modelred}
\end{align}
with the functional form of $M^{*}_{B,{\rm Q}}\,(z)$ and $\phi^{*}_{\rm Q}(z)$ as presented in Eqs.~(\ref{mstarz}) and (\ref{phiz}). The second Schechter function, defined with the faint characteristic luminosity $M_{\rm f}$ and the faint-end slope $\beta$, was needed to model the excess of faint quiescent galaxies \citep[e.g.][]{madgwick03,drory09,loveday12}. We tested several combinations of parameters and concluded that the luminosity function of quiescent galaxies is well described with these seven parameters, $\boldsymbol{\theta}^{\rm Q}_{\Phi} = [M^{0}_{B,{\rm Q}}, Q_{\rm Q}, \phi^0_{\rm Q}, P_{\rm Q}, \alpha_{\rm Q}, M_{\rm f}, \beta]$. Finally, a total of 12 parameters were needed to parametrise the joint quiescent and star-forming luminosity function. 

We note that the final model $\Phi_{\rm mod}$ was affected by the same $I_0 \leq 24$ selection than the observational data. With this approach, we used all the available information to compute a limited set of parameters, accounting for the correlations between variables and avoiding the completeness limit imposted by the probed effective volume.

\begin{figure*}[t]
\centering
\resizebox{0.49\hsize}{!}{\includegraphics{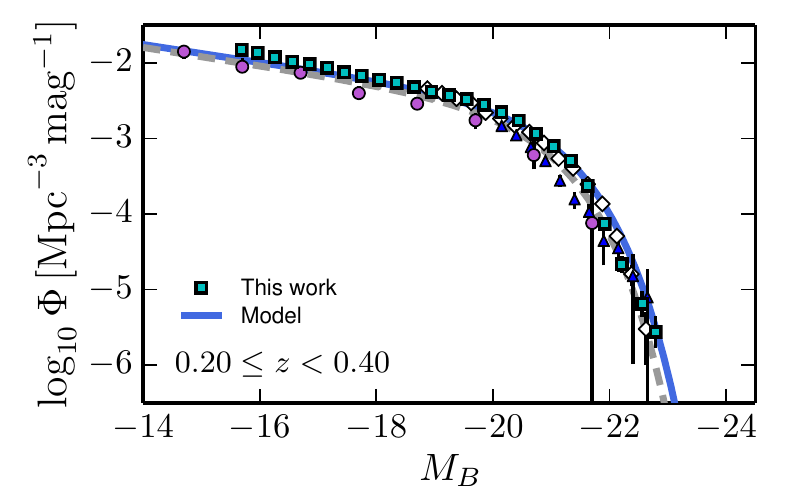}}
\resizebox{0.49\hsize}{!}{\includegraphics{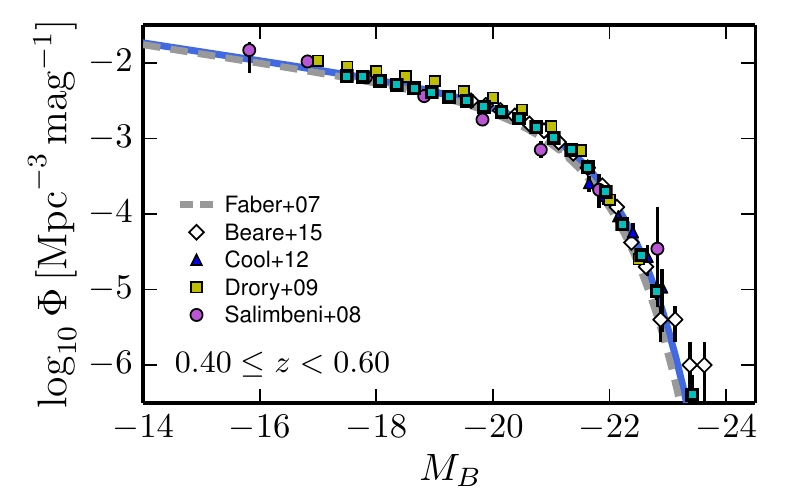}}\\
\resizebox{0.49\hsize}{!}{\includegraphics{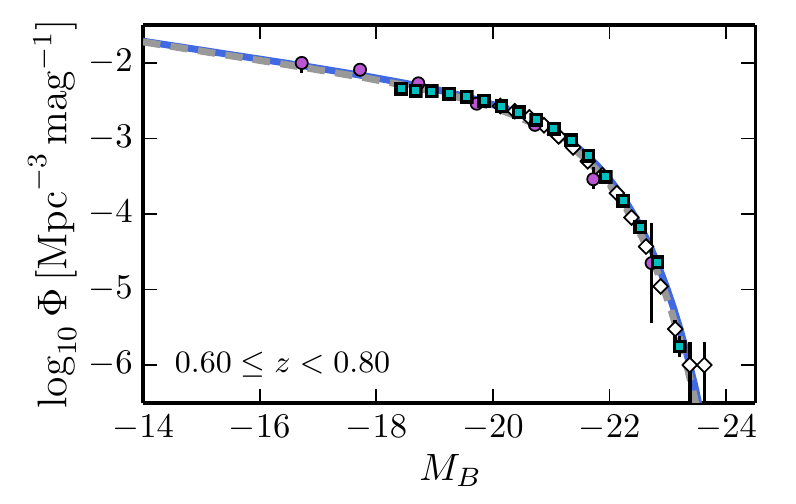}}
\resizebox{0.49\hsize}{!}{\includegraphics{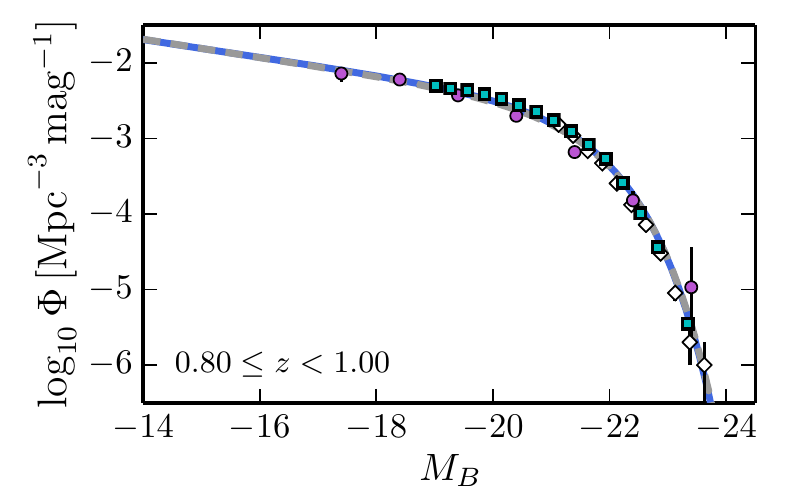}}\\
\caption{ALHAMBRA luminosity function of star-forming galaxies in four redshift bins (labelled in the panels). The blue squares are the observed luminosity functions and the blue solid lines the median model. The grey dashed lines are the best fitting from \citet{faber07}, including DEEP2, COMBO-17, 2dFGS \citep{madgwick02}, and SDSS \citep{bell03}. The purple dots are from \cite{salimbeni08} in GOODS-MUSIC, the blue triangles from \citet{cool12} in AGES, the green squares from \citet{drory09} in the COSMOS field, and the white diamonds from \citet{beare15} in B\"ootes.}
\label{lf_blue}
\end{figure*}

\begin{figure*}[t]
\centering
\resizebox{0.49\hsize}{!}{\includegraphics{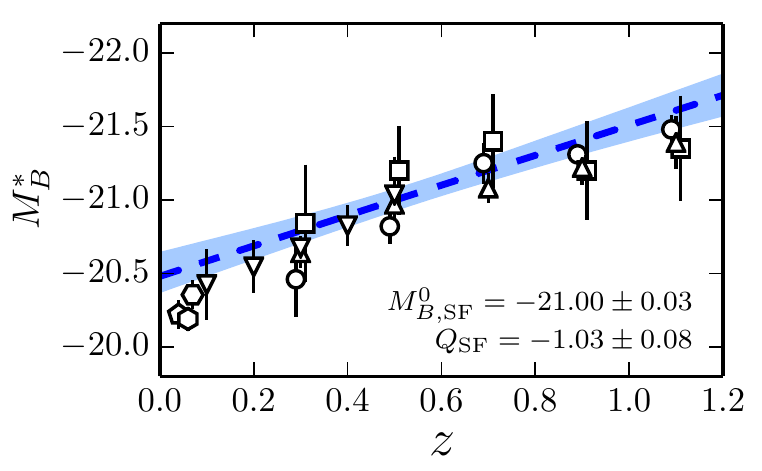}}
\resizebox{0.49\hsize}{!}{\includegraphics{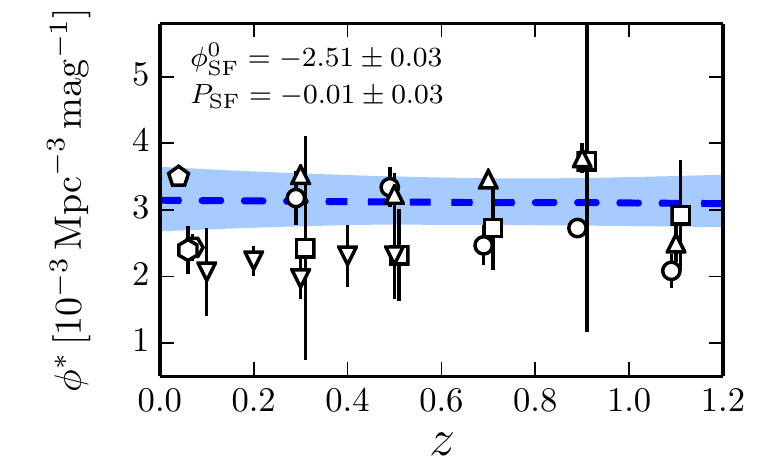}}\\
\resizebox{0.49\hsize}{!}{\includegraphics{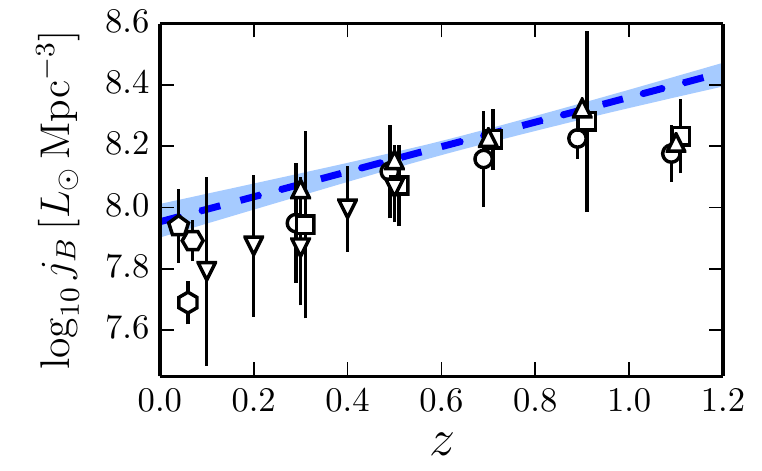}}
\resizebox{0.49\hsize}{!}{\includegraphics{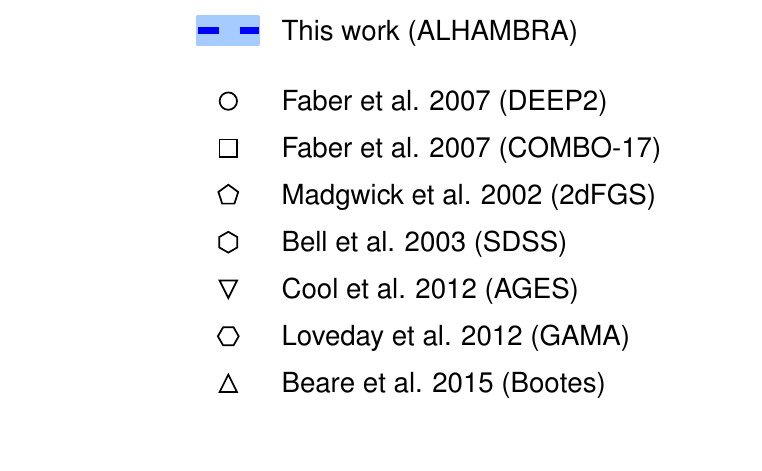}}
\caption{Redshift evolution for star-forming galaxies of $M_B^* \propto Q_{\rm SF}\,z$ ({\it top left panel}), $\phi^* \propto 10^{P_{\rm SF}\,z}$ ({\it top right panel}), and the $B-$band luminosity density $j_B$ ({\it bottom left panel}). In all panels the blue dashed line show the median model from ALHAMBRA data, with the coloured areas enclosing 95\% of the solutions. The median of the parameters with their associated 68\% ($1\sigma$) probability intervals are labelled in the panels. The other symbols, labelled in the {\it bottom right panel}, are from the literature. Their error bars mark 2$\sigma$ confidence intervals.}
\label{msphi_blue}
\end{figure*}

\subsubsection{Galaxy bias function model}\label{biasfit}
The galaxy bias function of the galaxy population $t$ is modelled as
\begin{equation}
b_{v,{\rm mod}}^{t}\,(z,M_B\,|\,\boldsymbol{\theta}_{b}^{t}) = A_t + B_t\,10^{-0.4[M_B - M^*_{B,t}(z)]},
\end{equation}
where the parameters $\boldsymbol{\theta}_{b}^{\rm Q} = [A_{\rm Q},B_{\rm Q}]$ and $\boldsymbol{\theta}_{b}^{\rm SF} = [A_{\rm SF},B_{\rm SF}]$ were estimated independently, as justified in Sect.~\ref{covabias}. In these cases, the $\chi^2$ function that we plugged into Eq.~(\ref{pbayes}) is
\begin{equation}
\chi^2\,(\tens{b}^{t}_v\,|\,\boldsymbol{\theta}_{b}^{t},\tens{\Sigma}_{b}^{t}) = [\tens{b}^{t}_v - \tens{b}^t_{v,{\rm mod}}]^{\rm T}\,[\tens{\Sigma}_{b}^{t}]^{-1}\,[\tens{b}^{t}_v - \tens{b}^t_{v,{\rm mod}}],\label{biasml}
\end{equation}
where $\tens{b}_{v,{\rm mod}}^{c}$ is the binned version of the bias function model defined with Eq.~(\ref{bvbin}). As in the luminosity function case, uninformative flat priors were assumed. We explored a linear dependence of the parameters $A_t$ and $B_t$ with redshift, but in all the cases such evolution was compatible with zero.

\begin{figure*}[!t]
\centering
\resizebox{0.49\hsize}{!}{\includegraphics{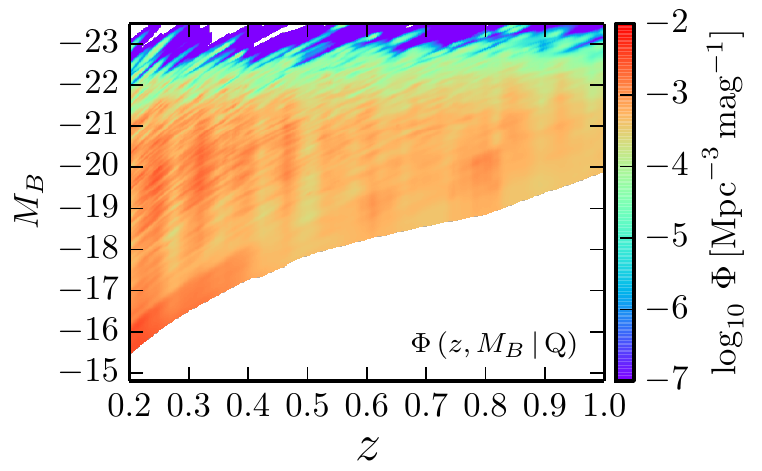}}
\resizebox{0.49\hsize}{!}{\includegraphics{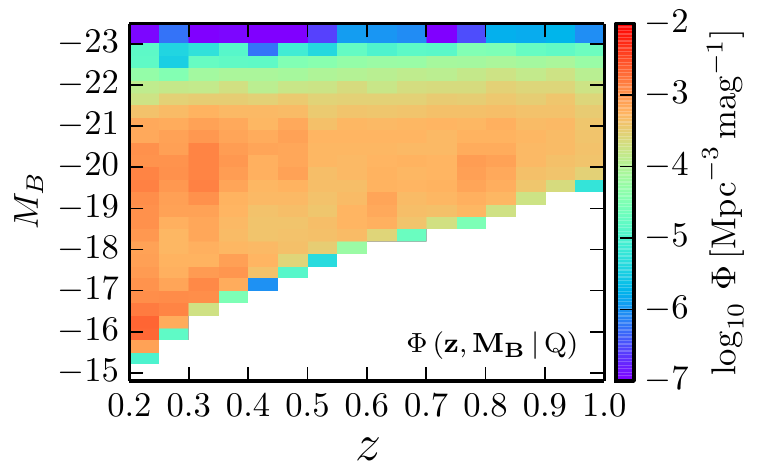}}\\
\resizebox{0.49\hsize}{!}{\includegraphics{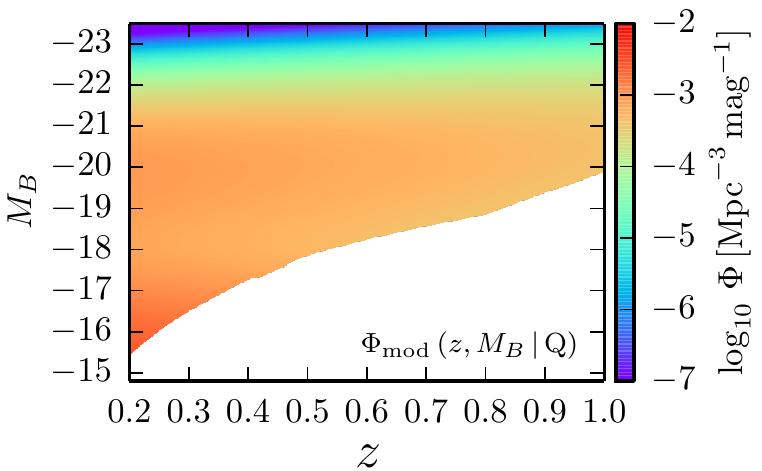}}
\resizebox{0.49\hsize}{!}{\includegraphics{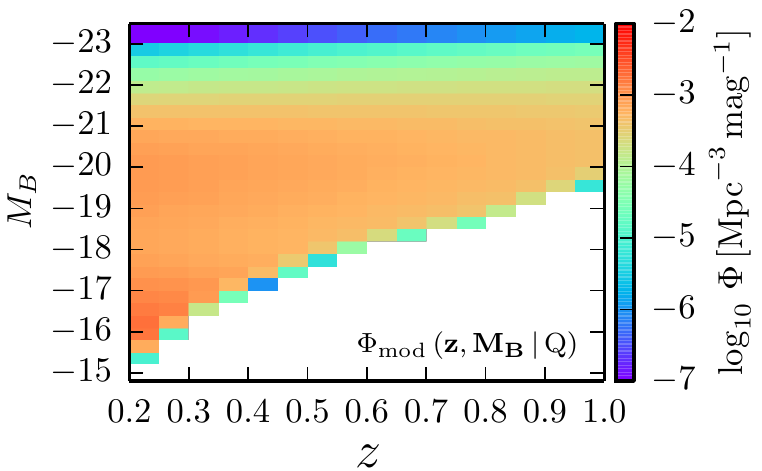}}
\caption{{\it Top panels}: Posterior luminosity function of quiescent galaxies in ALHAMBRA, $\Phi\,(z,M_B\,|\,{\rm Q})$, differential ({\it left}) and binned ({\it right}) version with $\Delta z = 0.05$ and $\Delta M_B = 0.3$. {\it Bottom panels}: Median luminosity function model for quiescent galaxies, differential ({\it left}) and binned ({\it right}) version. These luminosity functions are complete for $I_0 \leq 24$ galaxies.}
\label{lf2d_red}
\end{figure*}

\begin{figure*}[t]
\centering
\resizebox{0.49\hsize}{!}{\includegraphics{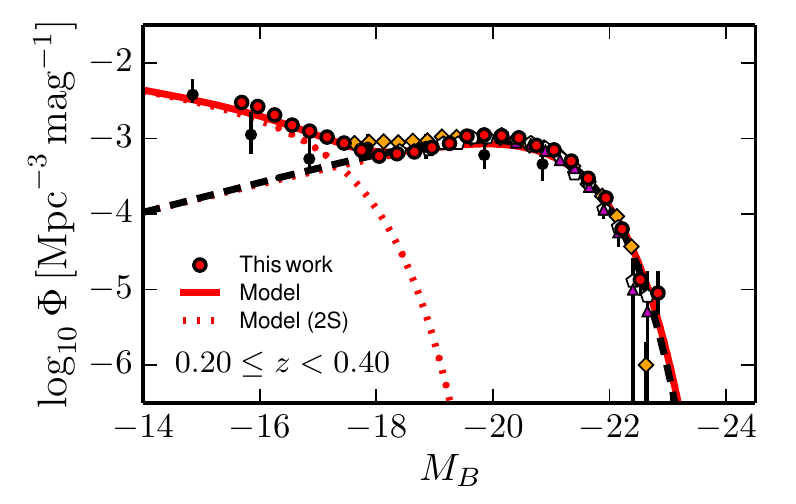}}
\resizebox{0.49\hsize}{!}{\includegraphics{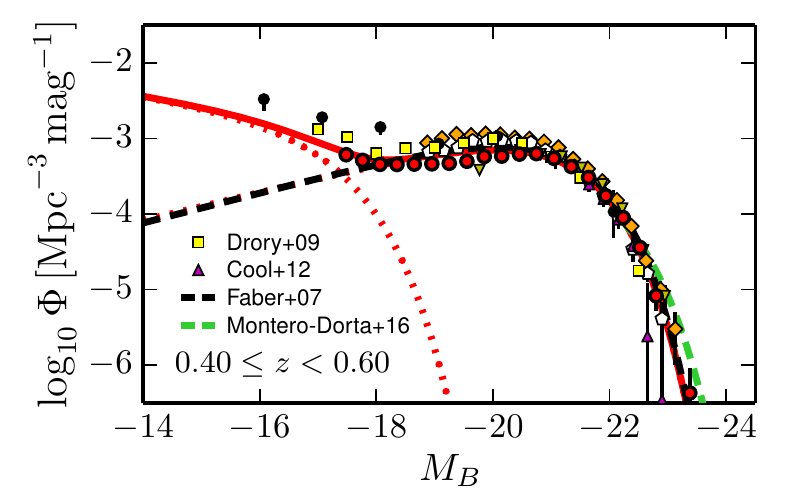}}\\
\resizebox{0.49\hsize}{!}{\includegraphics{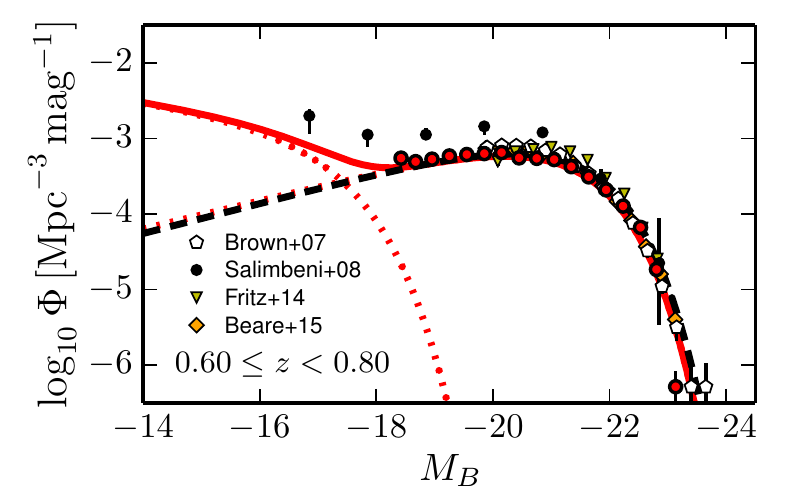}}
\resizebox{0.49\hsize}{!}{\includegraphics{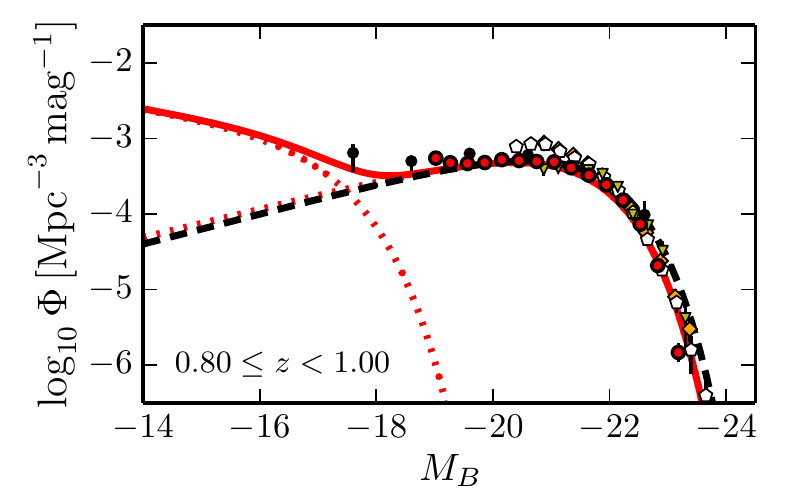}}\\
\caption{ALHAMBRA luminosity function of quiescent galaxies in four redshift bins (labelled in the panels). The red dots are the observed luminosity functions and the red solid line the median model. The dotted red lines show the bright and faint component of the median model. The dashed black lines are the best fitting from \citet{faber07}, including DEEP2, COMBO-17, 2dFGS \citep{madgwick02}, and SDSS \citep{bell03}. The dashed green line is from \citet{monterodorta16lf} in BOSS at $z = 0.55$. The white pentagons are from \citet{brown07} in the NDWFS, the black dots from \cite{salimbeni08} in GOODS-MUSIC, the purple triangles from \citet{cool12} in AGES, the yellow squares from \citet{drory09} in the COSMOS field, the inverted green triangles from \citet{fritz14} in VIPERS, and the orange diamonds from \citet{beare15} in B\"ootes.}
\label{lf_red}
\end{figure*}

\begin{figure*}[t]
\centering
\resizebox{0.49\hsize}{!}{\includegraphics{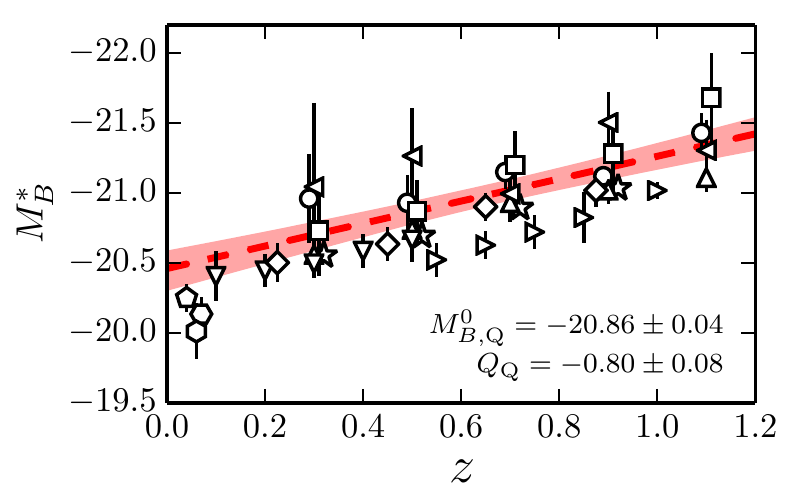}}
\resizebox{0.49\hsize}{!}{\includegraphics{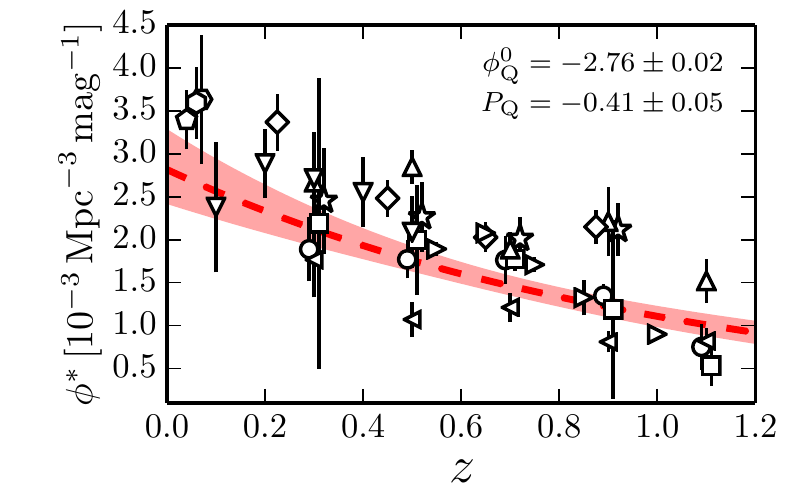}}\\
\resizebox{0.49\hsize}{!}{\includegraphics{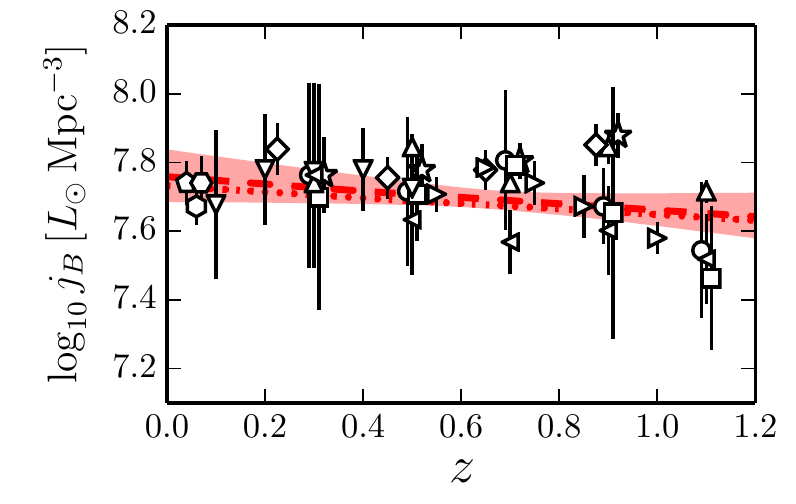}}
\resizebox{0.49\hsize}{!}{\includegraphics{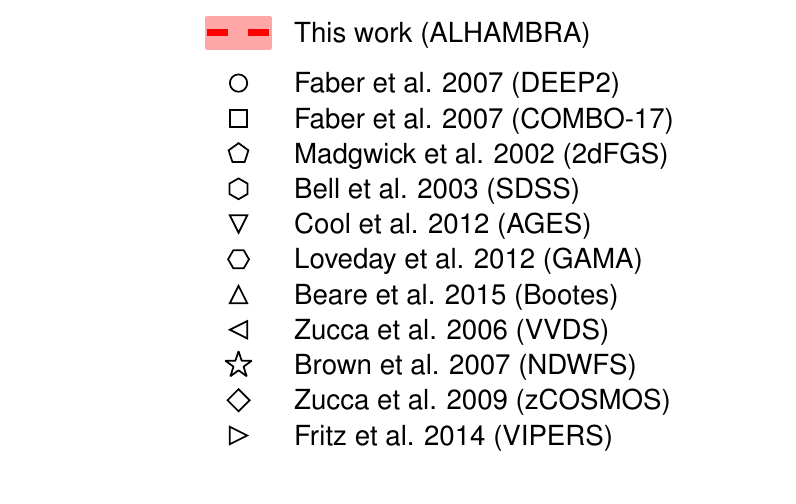}}
\caption{Redshift evolution for quiescent galaxies of $M_B^* \propto Q_{\rm Q}\,z$ ({\it top left panel}), $\phi^* \propto 10^{P_{\rm Q}\,z}$ ({\it top right panel}), and the $B-$band luminosity density $j_B$ ({\it bottom left panel}). In all panels the red dashed line show the median model from ALHAMBRA data, with the coloured areas enclosing 95\% of the solutions. The red dotted line in the {\it bottom left panel} shows the median model for the bright Schechter component alone. The median of the parameters with their associated 68\% (1$\sigma$) probability intervals are labelled in the panels. The other symbols, labelled in the {\it bottom right panel}, are from the literature. Their error bars mark 2$\sigma$ confidence intervals.}
\label{msphi_red}
\end{figure*}

\begin{table*}
\caption{ALHAMBRA luminosity function parameters.}
\label{param_tab}
\begin{center}
\begin{tabular}{lccccccc}
\hline\hline\noalign{\smallskip}
Galaxy type &  $M_B^{0}$ & $Q$ & $\phi^{0}$ & $P$ & $\alpha$ & $M_{\rm f}$ & $\beta$\\
\noalign{\smallskip}
\hline
\noalign{\smallskip}
Star-forming 	& $-21.00 \pm 0.03$ & $-1.03 \pm 0.08$ & $-2.51 \pm 0.03$ & $-0.01 \pm 0.03$ & $-1.29 \pm 0.02$ & $\cdots$ & $\cdots$\\
Quiescent  	& $-20.86 \pm 0.04$ & $-0.80 \pm 0.08$ & $-2.76 \pm 0.02$ & $-0.41 \pm 0.05$ & $-0.53 \pm 0.04$ & $-17.00 \pm 0.09$ & $-1.31 \pm 0.11$\\
\hline
\end{tabular}
\end{center}
\tablefoot{The quoted intervals represent 68\% ($1\sigma$) probability solutions.}
\end{table*}

\subsubsection{Joint modelling}\label{joint}
We note that the parametrization of the galaxy bias depends on $M^{*}_B$, and the luminosity function fitting depends on the galaxy bias because of the cosmic variance term in $\tens{\Sigma}_{\Phi}$. Thus, we performed an iterative fitting, starting with $b_{v,{\rm mod}} = 1$. We find that this process rapidly converges and just a few iterations were needed. We explored the posterior distribution from Eq.~(\ref{pbayes}) with the \texttt{emcee} \citep{emcee} code, a \texttt{Python} implementation of the affine-invariant ensemble sampler for Markov chain Monte Carlo (MCMC) proposed by \citet{goodman10}. The \texttt{emcee} code provides a collection of solutions in the parameter space, with the density of solutions being proportional to the posterior probability of the parameters. We obtained the more probable values of the parameters and their uncertainties as the median and the dispersion of the projected solutions.

Once the convergence was reached, we performed a total of ten iterations and, in each of them, we used a random solution of the set provided by \texttt{emcee} in the previous iteration to compute $\tens{b}_{v}^{t}$ and $\tens{\Sigma}_v$. This procedure ensures that the uncertainties in the fitting of the bias function were translated to the covariance matrix of the luminosity function, providing a representative mapping of the parameters space. The final collection of solutions was obtained as the combination of the ten iterations and our final set of parameters $\boldsymbol{\theta}_{\Phi}$, $\boldsymbol{\theta}_{b}^{\rm Q}$, and $\boldsymbol{\theta}_{b}^{\rm SF}$ describe the joint luminosity and galaxy bias functions consistently.

\subsubsection{Accounting for the prior uncertainty in the fitting process}\label{priorfit}
With the scheme presented in the previous sections, we have missing one key component in our final error budget. The estimation of the photometric redshift ${\rm PDF}\,(z,T)$ assumes a prior probability that is computed from the luminosity function \citep{benitez00}. Thus, the ALHAMBRA luminosity function could be biased towards the literature values used to compute the \texttt{BPZ2} prior probability. To trace the impact of the assumed prior in our results, we performed the joint modelling described in Sect.~\ref{joint} with three different priors: (i) the default \texttt{BPZ2} prior, (ii) a constant prior, and (iii) a cosmological volume prior. We checked that the priors (ii) and (iii) are extreme, and any realistic photometric redshift prior would be enclosed by them. The final parameters presented in Sect.~\ref{results} are the combination of the three prior solutions, and we discuss the impact of the prior on the fitted parameters in Sect.~\ref{priorparam}.

\section{ALHAMBRA $B-$band luminosity function}\label{results}
In this section we present the ALHAMBRA luminosity function of both star-forming (Sect.~\ref{lfblue}) and quiescent (Sect.~\ref{lfred}) galaxies, and their associated galaxy bias functions (Sect.~\ref{bias}). As a previous step, we explore the optimum binning in the luminosity and galaxy bias function estimation in Sect.~\ref{binning}.

\subsection{Optimum binning in ALHAMBRA}\label{binning}
The bin sizes $\Delta z$ and $\Delta M_B$ are fundamental in the estimation of $\tens{\Phi}$ and in the final fitting process. The probabilistic nature of the PDFs lead to correlations because each galaxy is spread over several adjacent bins. Even with the covariance matrix in hands (Sect.~\ref{covalf}), we have to ensure that the error budget of the binned luminosity function is understood and robustly estimated.

Several studies use mathematical arguments to define the optimum bin size \citep[e.g.][]{Shimazaki07}, but we used physical arguments thanks to the diagonal values (variances) of $\tens{\Sigma}_{\Phi}$ and $\tens{\Sigma}_v$, noted $\sigma_{\Phi}^2$ and $\sigma_v^2$. Both the total variance and the cosmic variance have to decrease if the volume probed by each bin (i.e. the bin size) increases. To study the impact of the assumed $\Delta z$ in the total variance, we measured the median variance of the bins in the redshift range $0.2 < z \leq 1$,
\begin{equation}
\langle \sigma \rangle = \langle \sigma_{\Phi}\,(0.2 \leq z_n < 1) \rangle.
\end{equation}
In the present exercise, the variances were estimated only at different redshift bins and all the luminosities and galaxy types were taken into account together. We present the variation of $\langle \sigma \rangle$ in the redshift range $0.2 \leq z < 1$ for $I_0 \leq 24$ galaxies in the top panel of Fig.~\ref{sig_fig}. We find that the total variance is constant, $\langle \sigma \rangle \sim 0.28$, for bin sizes smaller than the transition scale $\Delta z_{\rm T} \sim 0.02$, then the variance starts to decrease. The expected increase in the variance at $\Delta z < \Delta z_{\rm T}$ is spread in the non-diagonal terms of the covariance matrix, leaving constant the diagonal elements measured by $\sigma_{\Phi}$, and dominating the correlations between adjacent bins the error budget at small bin sizes. Interestingly, the measured transition scale is similar to the resolution of the photometric redshifts at $0.2 \leq z < 1$, $\langle \delta_z \rangle = 0.012\,(1+\langle z \rangle) = 0.019$ \citep{molino13}. We repeated this experiment but measuring the median cosmic variance. In this case, the cosmic variance increases as expected down to $\Delta z_v \sim 0.05$, then starts to unexpectedly decrease at smaller bin sizes. This implies that at smaller redshift bins our assumption of a log-normally distributed shot noise fails (Sect.~\ref{poisson}), and to measure reliable cosmic variances in ALHAMBRA we need $\Delta z \gtrsim 0.05$. Thus, we set $\Delta z = 0.05 \sim 2\langle \delta_z \rangle$ as our fiducial redshift bin in the study of the luminosity function as it ensures a well controlled error budget, both shot noise and cosmic variance. We test this conclusion by repeating the study at brighter magnitudes ($I_0 \leq 22.5$, bottom panel in Fig.~\ref{sig_fig}). We find that both $\Delta z_{\rm T}$ and $\Delta z_v$ are the same and only changes the normalisation of the curves, with $\langle \sigma \rangle \sim 0.43$ in the constant regime. This implies that our conclusions can be assumed independent of the $I-$band selection, reinforcing our interpretation of the trends.

Following the previous reasoning, we set the $B$-band bin size to $\Delta M_B = 0.3$, which is roughly twice the typical error in the $B-$band magnitude, $\sigma_{\texttt{BPZ}} \sim 0.12$, as estimated in Sect.~\ref{zmbpos}. To ensure reliable measurements in the bright end of the luminosity function, the brightest bin at every redshift is composed by those galaxies with $M_B \leq -23$. If brighter bins were used, the inverse of the covariance matrix became divergent because of the low statistics. At the faint end, only luminosity bins with $M_B \leq -15.5$ were taking into account and those with $V_{\rm eff}/\Delta V < 0.01$ were discarded. Finally, our binned luminosity function is a $2 \times 16 \times 26$ tensor, comprising two galaxy types, 16 redshift bins of $\Delta z = 0.05$, and 26 magnitude bins of $\Delta M_B = 0.3$. Taking into account the accessible volumes in the $z - M_B$ space, we had 586 data points to constrain 12 parameters.

In the galaxy bias analysis even larger volumes were needed. We explored several redshift and luminosity bins combinations, concluding that robust results were reached with three redshift bins, $0.2 \leq z < 0.65$, $0.65 \leq z < 0.85$, and $0.85 \leq z < 1$, and several luminosity bins in steps of $0.5$ magnitudes with the brightest one comprising $M_{B} - M_B^{*} \leq 0$ galaxies. We checked that with smaller bins the measured cosmic variance starts to decrease non-physically, reflecting the trend observed in Fig.~\ref{sig_fig}. In this case we have 15 points to estimate the bias function of star-forming galaxies, and 14 points in the quiescent galaxies case. This limited number of data points compared with the luminosity function case reflects the statistical difficulties of study the intrinsic dispersion of the galaxy distribution.

\subsection{ALHAMBRA luminosity function}
In this section we present the ALHAMBRA luminosity function computed with the fiducial \texttt{BPZ2} photometric redshift prior. We do not present the luminosity functions computed with the constant and volume prior (Sect.~\ref{priorfit}) for the sake of clarity, but discuss the impact of the prior in Sect.~\ref{priorparam}. These three luminosity functions are made public together with their covariance matrices for three redshift bin sizes ($\Delta z = 0.05, 0.1, 0.2$) at the PROFUSE web page. We stress that the modelling presented in this section includes the photometric redshift prior uncertainties. The derived parameters $\boldsymbol{\theta}_{\Phi}$, $\boldsymbol{\theta}_{b}^{\rm Q}$, and $\boldsymbol{\theta}_{b}^{\rm SF}$ are summarised in Tables~\ref{param_tab} and \ref{bias_tab}. We focus on star-forming galaxies in Sect.~\ref{lfblue} and on quiescent galaxies in Sect.~\ref{lfred}. The effective number of galaxies under study, computed as the integral of the $z - M_B$ PDFs, is 80464 star-forming and 16125 quiescent. We discuss the ALHAMBRA quiescent fraction in Sect.~\ref{fred}.

\subsubsection{ALHAMBRA luminosity function of star-forming galaxies}\label{lfblue}
The estimated ALHAMBRA luminosity function of star-forming galaxies, $\Phi\,(z,M_B\,|\,{\rm SF})$, is presented in Fig.~\ref{lf2d_blue}. We show both the differential and binned versions, and also the median model from Eq.~(\ref{modelblue}). The differential version of the luminosity function presents over-dense strips in redshift space, reflecting the presence of cosmic structures. Also exists strips as consequence of the known $z-M_B$ correlation (Fig.~\ref{pdfzmb}). All these structures vanishes in the binned version of the luminosity function, but the incompleteness at the faint end due to volume effects is evident. We stress that such volume effects are taking into account in the modelled luminosity function, that reproduce not only the observed trends, but also the apparent lack of sources in the fainter bins.

To facilitate the comparison with previous studies in the literature, we present the ALHAMBRA luminosity function in four redshift bins with $\Delta z = 0.2$ in Fig.~\ref{lf_blue} and Appendix~\ref{tables}. In both cases the values at the faint end were computed using the effective volume $V_{\rm eff}$ instead of $\Delta V$, but we recall that in the fitting process such correction was not performed. We also show the luminosity function values from previous work in Fig.~\ref{lf_blue}, including among others the results from COMBO-17 \citep{combo17}, 2dFGS \citep{2dfgs}, AGES \citep{ages}, and GOODS-MUSIC \citep{goodsmusic} surveys. We find that the ALHAMBRA luminosity function of star-forming galaxies agrees with previous results in the literature. We stress the agreement with the results from \citet{faber07}, based on several surveys (DEEP2, COMBO-17, 2dFGS, SDSS), with the 8.26 deg$^2$ studied by \citet{beare15} in the B\"ootes field, and with the 1.5 mag deeper results of \citet{salimbeni08} in GOODS-MUSIC. We find a lower density of star-forming galaxies in AGES at $z = 0.3$ and a larger density in COSMOS at $z = 0.5$ with respect to ALHAMBRA. These fluctuations are probably related with the cosmic variance affecting these surveys.

We present the redshift evolution of $M_{B, {\rm SF}}^{*}$ and $\phi^{*}_{\rm SF}$ in Fig.~\ref{msphi_blue}. We show the median model from the iterative process described in Sect.~\ref{fitting}, including 95\% probability solutions. We find that the star-forming populations is well described by a redshift-evolving Schechter function with $Q_{\rm SF} = -1.03 \pm 0.08$, $P_{\rm SF} = -0.01 \pm 0.03$, and $\alpha_{\rm SF} = -1.29 \pm 0.02$. The quoted intervals represent 68\% ($1\sigma$) probability solutions. The derived evolution of $M_{B, {\rm SF}}^{*}$ is compatible with previous work, that covers the range $Q_{\rm SF} \in (-0.6,-1.3)$. The redshift evolution of $\phi^{*}_{\rm SF}$ is compatible with zero, implying a roughly constant value. Our estimation is in good agreement with the results from the 2dFGS \citep{madgwick02}, B\"ootes \citep{beare15}, and DEEP2 \citep{faber07}, but seems too low when compared with AGES \citep{cool12} and \citet{loveday12} in GAMA \citep{gama}, that suggest a positive value of $P_{\rm SF}$. The derived value of $\alpha_{\rm SF} = -1.29$ agrees with previous work, that find a value in the range $\alpha_{\rm SF} \in (-1.1,-1.4)$. We note that because of our fitting process, the value of the faint-end slope is constrained by those redshifts with relevant information at faint magnitudes, and such information is propagated to the higher redshift bins with limited coverage of the faint end.

We conclude that the ALHAMBRA luminosity function of star-forming galaxies agrees with previous results in the literature, and that provides a consistent evolution of both $M_{B, {\rm SF}}^{*}$ and $\phi^{*}_{\rm SF}$. We explore the implications for the luminosity density evolution up to $z \sim 1$ in Sect.~\ref{jbsec}.

\begin{figure}[t]
\centering
\resizebox{\hsize}{!}{\includegraphics{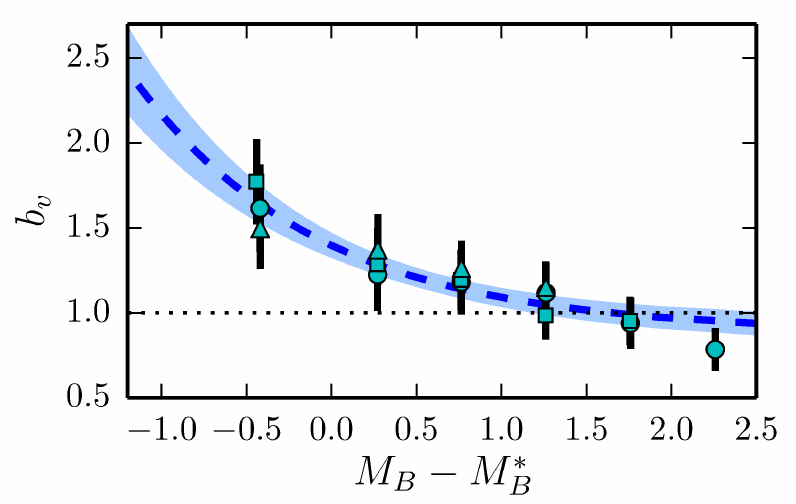}}\\
\resizebox{\hsize}{!}{\includegraphics{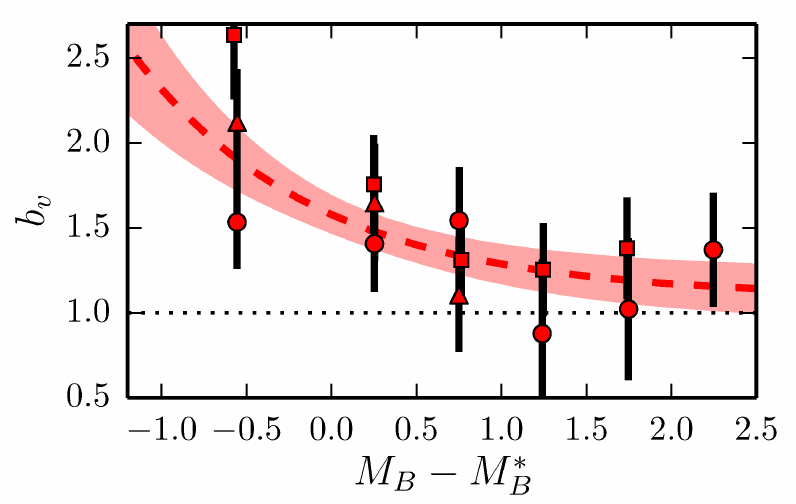}}
\caption{Bias function of star-forming ({\it top}) and quiescent ({\it bottom}) galaxies as a function of $M_B - M_B^{*}$ for three different redshift bins, $0.2 \leq z < 0.65$ (dots), $0.65 \leq z < 0.85$ (squares), and $0.85 \leq z < 1$ (triangles). The dashed lines show the median model to the ALHAMBRA data, with the coloured areas enclosing 68\% of the solutions. The dotted lines mark a galaxy bias $b_{v} = 1$.}
\label{biasal}
\end{figure}

\subsubsection{ALHAMBRA luminosity function of quiescent galaxies}\label{lfred}
The estimated ALHAMBRA luminosity function of quiescent galaxies, $\Phi\,(z,M_B\,|\,{\rm Q})$, is presented in Fig.~\ref{lf2d_red}, both the differential and binned versions. We also show the obtained median model from Eq.~(\ref{modelred}). As in the star-forming galaxies case, the differential version of the luminosity function presents structures in the $z-M_B$ space that vanishes in the binned version.

We present the ALHAMBRA luminosity function in four redshift bins with $\Delta z = 0.2$ in Fig.~\ref{lf_red} and Appendix~\ref{tables}. We also show the results from the literature in Fig.~\ref{lf_red}. Focusing on the bright end, which is covered by several previous studies, the ALHAMBRA luminosity function of quiescent galaxies agrees with \citet{faber07}, \citet{beare15}, \citet{cool12}, \citet{fritz14} in VIPERS, \citet{brown07} in the NDWFS survey, and \citet{monterodorta16lf} in BOSS at $z = 0.55$. We stress the apparent lack of ALHAMBRA galaxies at $z \sim 0.5$. This lower density is also found in the stellar mass analysis of D\'{\i}az-Garc\'{\i}a et al. (in prep.) using \texttt{MUFFIT} and different photometric redshift codes, so we conclude that ALHAMBRA is still affected by large scale structures.

We clearly find the up-turn of the luminosity function at magnitudes $M_B \gtrsim -18$, in agreement with the results from \citet{madgwick02} and \citet{loveday12} in the local Universe, \citet{drory09} at $z \sim 0.5$ in the COSMOS field, and \citet{salimbeni08} at $z \geq 0.4$ in GOODS-MUSIC. In the two last cases, we observe not only the same trends, but also the expected values from our median model are consistent with their observations. The comparison with the local Universe studies reveals a possible evolution of the luminosity $M_{\rm f}$, since the upturn appears at lower magnitudes than we observe ($M_B \sim -17$). The redshift range covered by the ALHAMBRA faint sources is limited ($0.2 \leq z \lesssim 0.5$), and such redshift evolution could be missed. The J-PAS survey will close the gap between local and higher redshift studies, providing a definitive picture to the possible redshift evolution of the faint, quiescent population.  

We present the evolution of $M_{B,{\rm Q}}^{*}$ and $\phi^{*}_{\rm Q}$ in Fig.~\ref{msphi_red}. We show the median model from the iterative process described in Sect.~\ref{fitting}, including 95\% probability solutions. We find that quiescent galaxies are well described by a redshift-evolving Schechter function with $Q_{\rm Q} = -0.80 \pm 0.08$, $P_{\rm Q} = -0.41 \pm 0.05$, and $\alpha_{\rm Q} = -0.53 \pm 0.04$. The second Schechter function that models the observed faint-end upturn has a slope $\beta = -1.31 \pm 0.11$. The quoted intervals represent 68\% ($1\sigma$) probability solutions. The derived evolution of $M_{B,{\rm Q}}^{*}$ agrees with previous studies, including among others \citet{zucca06} in the VVDS \citep{lefevre05,vvdsud} and \citet{zucca09} in zCOSMOS \citep{zcosmos10k}, that cover the range $Q_{\rm Q} \in (-0.4,-1.2)$. However, the ALHAMBRA values are significantly brighter than those in VIPERS \citep{fritz14}. We argue that this difference is due to their estimated value of $\alpha_{\rm Q}$, which is always positive and deviates from our preferred value and previous work, that usually find $\alpha_{\rm Q} \sim -0.5$. As stress by \citet{beare15}, $\alpha$ is correlated with $M_B^*$, and the more negative is $\alpha$, the brighter is $M_B^{*}$. Because the measured luminosity functions agree well (Fig.~\ref{lf_red}), the difference in $\alpha_{\rm Q}$ might explain the apparent difference on $M_B^{*}$. We also find that $\phi^{*}_{\rm Q}$ decreases with redshift, again in agreement with the values from the literature, that cover the range $P_{\rm Q} \in (-0.1,-0.5)$. Finally, we checked that our second faint-end slope $\beta = -1.3$ is consistent with previous results: \citet{loveday12} estimate $\beta = -1.6 \pm 0.3$ in the GAMA survey at $z \sim 0.1$, and \citet{salimbeni08} found $\beta = -1.8 \pm 0.2$ at $z \geq 0.4$.

As for the star-forming population, we conclude that the ALHAMBRA luminosity function of quiescent galaxies agrees with the literature and that provides a consistent evolution of both $M_{B,{\rm Q}}^{*}$ and $\phi^{*}_{\rm Q}$. We explore the implications for the luminosity density evolution up to $z \sim 1$ in Sect.~\ref{jbsec}.

\subsection{Galaxy bias functions}\label{bias}
In this section we present the estimated galaxy bias functions, both for quiescent and star-forming galaxies. The measured bias values are compiled in Appendix~\ref{tables}. Both galaxy bias functions are shown in Fig.~\ref{biasal}. The different selections and assumptions make difficult the quantitative comparison with previous work, so we focus in the qualitative analysis of our results.

We find that the galaxy bias $b_{v}$ of both star-forming and quiescent galaxies increases with the $B-$band luminosity, a well known trend \citep[e.g.][]{norberg01,arnaltemur14}. The slopes $B_{\rm SF} = 0.51 \pm 0.10$ and $B_{\rm Q} = 0.49 \pm 0.16$ are compatible within their uncertainties, and also with the slope $B = 0.5 \pm0.24$ found by \citet{skibba14} in a similar redshift range. However, these slopes are larger than the local universe values, $B \sim 0.2$, based on clustering analysis \citep{norberg01,zehavi11}.

Regarding the normalization, we find that the bias of quiescent galaxies $A_{\rm Q} = 1.09 \pm 0.15$ is larger than the bias of star-forming galaxies $A_{\rm SF} = 0.89 \pm 0.07$. The inferred relative bias between the quiescent and star-forming populations is $b_{\rm rel} = A_{\rm Q} / A_{\rm SF} = 1.2 \pm 0.2$, in agreement with previous studies in ALHAMBRA \citep{clsj15bcosvar,lluis16} and other surveys \citep{madgwick03,meneux06,coil08,delatorre11,skibba14}. 

The statistical significance of our galaxy bias results is limited by the number of accessible sub-fields. The large area of the J-PAS survey will permit the definition of several hundred sub-fields, greatly improving the galaxy bias function analysis presented in the present paper.

\section{Discussion}\label{discussion}

\subsection{Evolution of the luminosity density}\label{jbsec}
The $B-$band luminosity density $j_B$ express the total amount of light emitted by galaxies in the $B$ band per unit volume, and it provides insights about the physical processes involved in the evolution of star-forming and quiescent galaxies with cosmic time. The luminosity density of a single Schechter luminosity function is 
\begin{align}
j_B\,(z) = \int & L_B\,\Phi\,(z,L_B)\,{\rm d}L_B =\nonumber\\ 
& \phi^{*}(z)\,10^{0.4[M_{B_{\odot}} - M_B^*(z)]}\,\Gamma(\alpha + 2)\,\,\,\,{\rm [}L_{\odot}\,{\rm Mpc}^{-3}{\rm ]},
\end{align}
where $M_{B_{\odot}} = 5.38$ mag is the $B$-band absolute magnitude of the Sun \citep{binney98}, and $\Gamma$ is the Gamma function.

We present the luminosity density of star-forming galaxies in the bottom left panel of Fig.~\ref{msphi_blue}. In agreement with previous results in the literature, we find that $j_B$ decreases by a factor of $2.55 \pm 0.14$ since $z = 1$. That reflects the descent in the star formation rate density of the Universe \citep[e.g.][]{hopkins06,cucciati12,sobral13}, that translates to a lower production of $B-$band photons by newly formed, massive stars with time. 

In the case of quiescent galaxies, we compute the luminosity density of the bright component and the faint component separately. Both components are presented in the bottom left panel of Fig.~\ref{msphi_red}. Regarding the total luminosity density in quiescent galaxies, we find a mild increase by a factor $1.25 \pm 0.16$ since $z = 1$. However, a passively evolving population produces less $B-$band photons as time goes by, and the arrival of new stars to the quiescent population is therefore needed to explain the observed increase of $j_B$ with cosmic time \citep{bell04,faber07,brown07}. The contribution of the faint quiescent population to the luminosity density increases from $3 \pm 1$\% at $z = 1$ to $6 \pm 1$\% at $z = 0$, implying a larger arrival rate of faint galaxies to the red sequence with respect to the bright population. The emergence of the faint quiescent population has been related with environmental processes \citep[e.g.][]{peng10}. We will study different environments in future work to obtain further clues about this topic.

\begin{table}
\caption{ALHAMBRA galaxy bias functions parameters.}
\label{bias_tab}
\begin{center}
\begin{tabular}{lcc}
\hline\hline\noalign{\smallskip}
Galaxy type & $A$ & $B$\\
\noalign{\smallskip}
\hline
\noalign{\smallskip}
Star-forming	& $0.89 \pm 0.07$ & $0.51 \pm 0.10$\\
Quiescent 	& $1.09 \pm 0.15$ & $0.49 \pm 0.16$\\
\hline
\end{tabular}
\end{center}
\tablefoot{The quoted intervals represent 68\% ($1\sigma$) probability solutions.}
\end{table}

\subsection{Quiescent fraction in ALHAMBRA}\label{fred}
\begin{figure*}[t]
\centering
\resizebox{0.49\hsize}{!}{\includegraphics{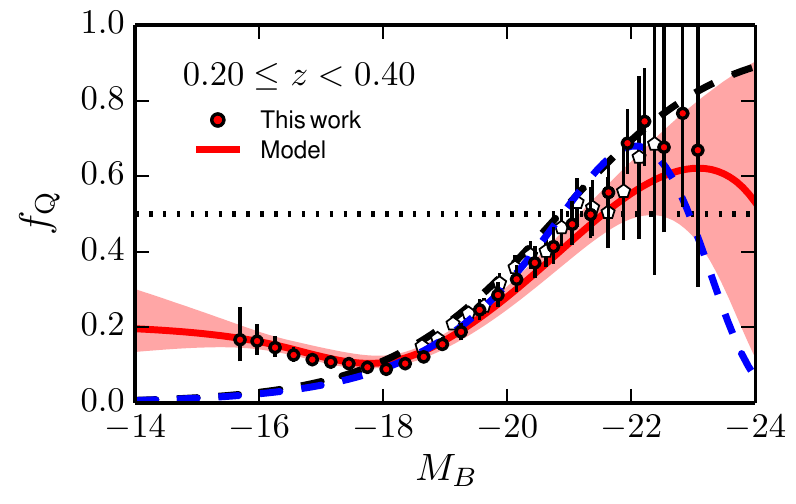}}
\resizebox{0.49\hsize}{!}{\includegraphics{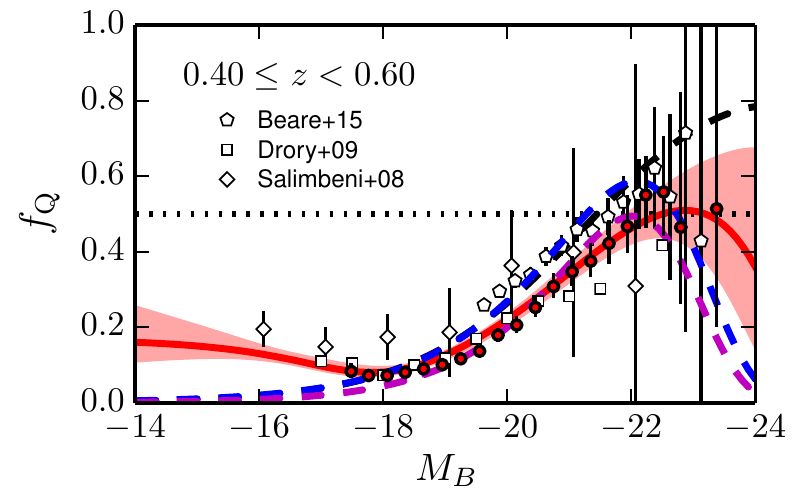}}\\
\resizebox{0.49\hsize}{!}{\includegraphics{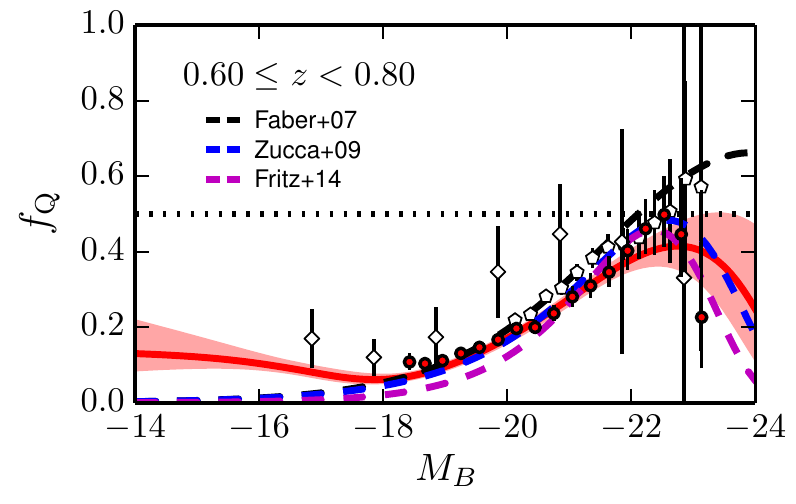}}
\resizebox{0.49\hsize}{!}{\includegraphics{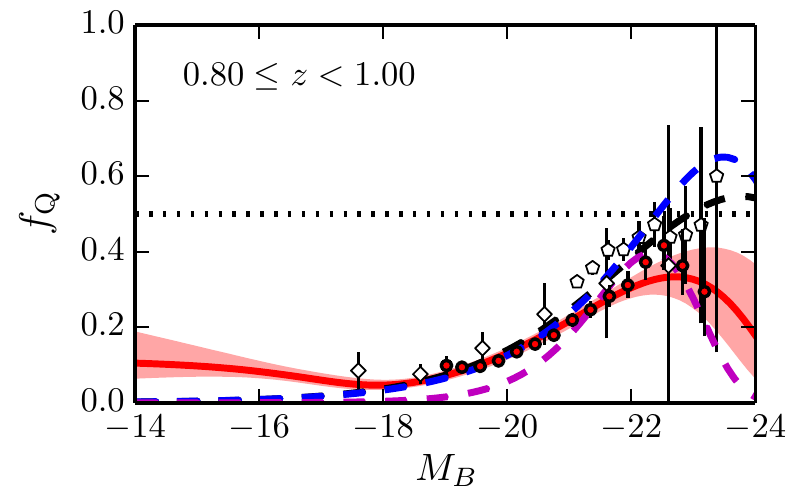}}\\
\caption{ALHAMBRA quiescent fraction $f_{\rm Q}$ in four redshift bins (labelled in the panels). The red dots are the observed $f_{\rm Q}$. The red solid line is the quiescent faction median model, and the coloured area its 95\% confidence range. The other symbols, labelled in the panels, are from the literature: the dashed lines are from the Schechter best fit of \citet{faber07}, \citet{zucca09}, and \citet{fritz14}, the white pentagons are from \citet{brown07}, the squares from \citet{drory09}, and the diamonds from \cite{salimbeni08}. The error bars mark $2\sigma$ confidence intervals. The dotted horizontal line marks $f_{\rm Q} = 0.5$.}
\label{fred_fig}
\end{figure*}

We complete the discussion about the ALHAMBRA luminosity function by studying the quiescent fraction as a function of redshift and $B-$band absolute magnitude. We estimate the quiescent fraction as
\begin{equation}
f_{\rm Q}\,({\bf z}, {\bf M_B}) = \frac{\Phi\,({\bf z}, {\bf M_B}\,|\,{\rm Q})}{\Phi\,({\bf z}, {\bf M_B}\,|\,{\rm Q}) + \Phi\,({\bf z}, {\bf M_B}\,|\,{\rm SF})}.
\end{equation}
We present the quiescent fraction in ALHAMBRA for $\Delta z = 0.2$ bins in Fig.~\ref{fred_fig}. We define three luminosity ranges: the faint regime at $M_B \geq -18$, the intermediate regime at $-18 \leq M_B \leq -21.5$, and the bright regime at $M_B \leq -21.5$. We find that, at any redshift range, the quiescent fraction has a minimum at $M_B \sim -18$ and increases towards brighter and fainter magnitudes. We also notice a maximum in $f_{\rm Q}$ in the bright regime, but the uncertainties in the measurements are large at these magnitudes and a steady increase is also compatible with the data. Regarding redshift evolution, the quiescent fraction decreases from $z = 0.3$ to $z = 0.9$ at any magnitude by a factor of $\sim 2$ on average. The quiescent fraction excess 50\% for $M_B \lesssim M_B^{*}$ galaxies at $z \lesssim 0.4$, and star-forming galaxies are always more numerous at $M_B \gtrsim M_B^{*}$, reaching a constant quiescent fraction in the faint regime. 

We also present the quiescent faction from previous work in Fig.~\ref{fred_fig}. The ALHAMBRA quiescent fraction is in good agreement with previous studies \citep{faber07,salimbeni08,zucca09,beare15,fritz14} at the intermediate regime, showing similar trends and values. The main discrepancy is with VIPERS \citep{fritz14} at $z \sim 0.9$, reflecting their positive $\alpha_{\rm Q}$ value (see Sect.~\ref{lfred}, for details). The faint regime is well covered by \citet{salimbeni08}, but the Schechter function extrapolation from other studies provides a $f_{\rm Q}$ that tends to zero. This is due to the missing faint quiescent population that it is only traced by \citet{salimbeni08} and \citet{drory09}, and that provide quiescent fractions consistent with ALHAMBRA. Finally, the bright regime is dominated by the uncertainties in the measurements. For example, the quiescent fraction of $M_B = -23$ galaxies at $z = 0.3$ covers the range $f_{\rm Q} \in (0.3,0.8)$. Future large-area surveys such as J-PAS will provide enough statistics to accurately constraint the bright end of the luminosity function and to test possible deviations from the widely used Schechter function \citep[e.g.][]{tempel09,bernardi10}.

\begin{figure}[t]
\centering
\resizebox{\hsize}{!}{\includegraphics{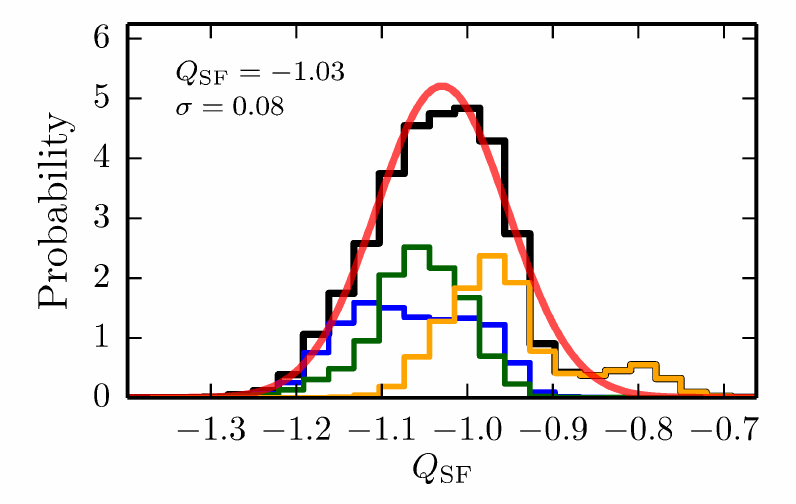}}\\
\resizebox{\hsize}{!}{\includegraphics{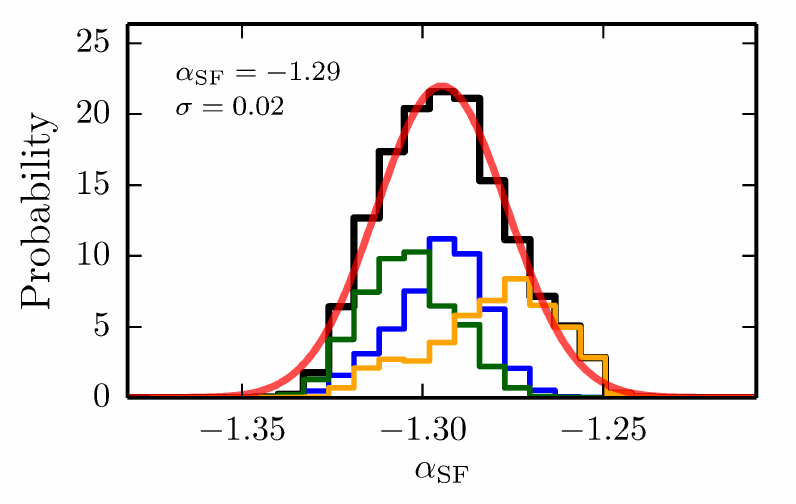}}\\
\resizebox{\hsize}{!}{\includegraphics{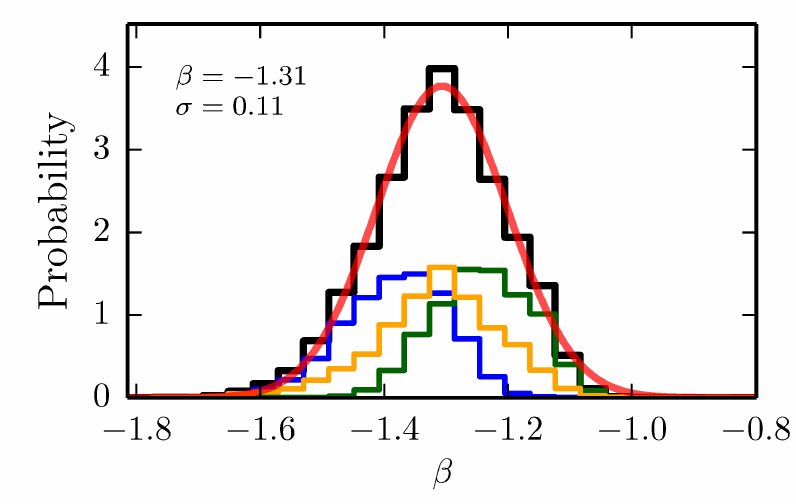}}
\caption{Impact of the photometric redshift prior in the estimated parameters. We present the estimated probabilities of $Q_{\rm SF}$ ({\it upper panel}), $\alpha_{\rm SF}$ ({\it central panel}), and $\beta$ ({\it lower panel}). The coloured histograms are the individual results obtained with \texttt{BPZ2} prior (blue), constant prior (green), and cosmic volume prior (orange). The combined final probability is the black histogram, with the best Gaussian fit to the distribution in solid red. The parameters of this Gaussian fit are labelled in the panels.}
\label{priorfig}
\end{figure}

\subsection{Impact of the prior in the luminosity function parameters}\label{priorparam}
We explore the impact of the prior in the final estimation of the luminosity function parameters in this section. We find that the three assumed priors (the fiducial \texttt{BPZ2} prior, the constant prior, and the cosmic volume prior) provide compatible luminosity functions at 2$\sigma$ level at any redshift and luminosity bin. The discrepancies exceed the 1$\sigma$ level only at $z \lesssim 0.5$ for $M_B \lesssim -18.5$ galaxies. This was expected because the importance of the prior increases as the signal--to--noise of the sources decreases (there are less information in the observed ALHAMBRA filters), and as the redshift of the sources decreases (the lower redshift solutions that are compatible within the data uncertainties are penalised by the available cosmological volume). We conclude that the prior uncertainty have a limited impact in our final luminosity functions.

Regarding the impact in the final fitting values, the individual parameters from the three priors are always compatible at 2$\sigma$ level, reflecting the observed differences in the measured luminosity functions. The impact of the prior is better illustrated in Fig.~\ref{priorfig}, where the contribution the three priors to the final solution of the parameters $Q_{\rm SF}$, $\alpha_{\rm SF}$, and $\beta$, is presented. The combined distribution is broader than the individual ones in any case, increasing the dispersion up to $\sim60$\%. For the parameters in Fig.~\ref{priorfig}, the dispersion increase is $50$\% ($Q_{\rm SF}$), $40$\% $(\alpha_{\rm SF}$), and $20$\% ($\beta$). We conclude that the assumed prior has a limited impact in our results, but it is a relevant source of uncertainty that should be included in the final error budget.

Finally, we study the impact of the quiescent or star-forming probability for red templates derived in Sect.~\ref{colpos}. We find that the faint-end of the red luminosity function is stepper than the quiescent one, with $\beta_{\rm red} = -1.6$ in comparison with $\beta_{\rm Q} = -1.3$. This difference remarks the importance of the contamination by dusty star-forming galaxies and the proper definition of the samples under study, as stress by \citet{taylor15}. The rest of parameters are less affected by dusty galaxies, with the red values compatible at $2\sigma$ level with the quiescent ones.

\begin{figure}[t]
\centering
\resizebox{\hsize}{!}{\includegraphics{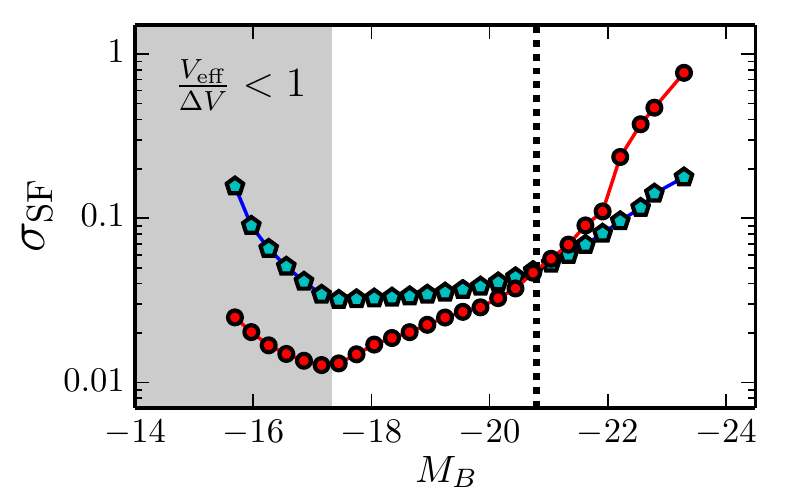}}\\
\resizebox{\hsize}{!}{\includegraphics{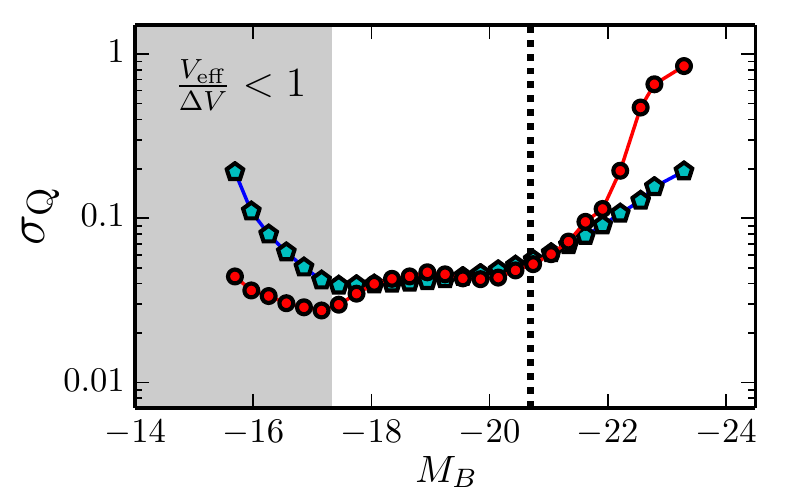}}
\caption{Relative variance (diagonal terms) of the luminosity function covariance matrix at $0.2 \leq z < 0.4$ as a function of $M_{B}$ for star-forming ({\it top panel}) and quiescent ({\it bottom panel}) galaxies. Red dots mark the shot noise contribution, and cyan pentagons the cosmic variance one. The grey areas in both panels show the magnitudes affected by the $I_0 \leq 24$ selection, where $V_{\rm eff}/\Delta V < 1$. The dotted vertical lines mark $M_B^{*}$ at $z = 0.3$ for reference.}
\label{diagfig}
\end{figure}

\subsection{Properties of the covariance matrix}\label{covdis}
Here we discuss in details the properties of the luminosity function covariance matrix computed in Sect.~\ref{covalf}. We start presenting the contribution of the shot noise and cosmic variance to the relative variance $\sigma$ of the data. Such variance is computed as the square root of the $\tens{\Sigma}_{\rm P}$ and $\tens{\Sigma}_v$ diagonal terms (see also Sect.~\ref{binning}), and represents the uncertainty reported in Appendix~\ref{tables} and shown in Figs.~\ref{lf_blue} and \ref{lf_red}. We show the variance of star-forming ($\sigma_{\rm SF}$) and quiescent ($\sigma_{\rm Q}$) galaxies as a function of the $B-$band absolute magnitude at $0.2 \leq z < 0.4$ in Fig.~\ref{diagfig}. We find that the shot noise typically dominates the error budget at $M_B \lesssim M_B^{*}$, with the cosmic variance being larger at lower luminosities. Both variances increase with luminosity because of the lower number density (shot noise) and the higher galaxy bias (cosmic variance), respectively, and at the faintest magnitudes because the probed cosmic volume is affected by the $I_{0}$ selection. These trends are in agreement with the theoretical expectations of \citet{smith12} and are also present in the other redshift ranges under study.

\begin{figure*}
\sidecaption
  \includegraphics[width=12cm]{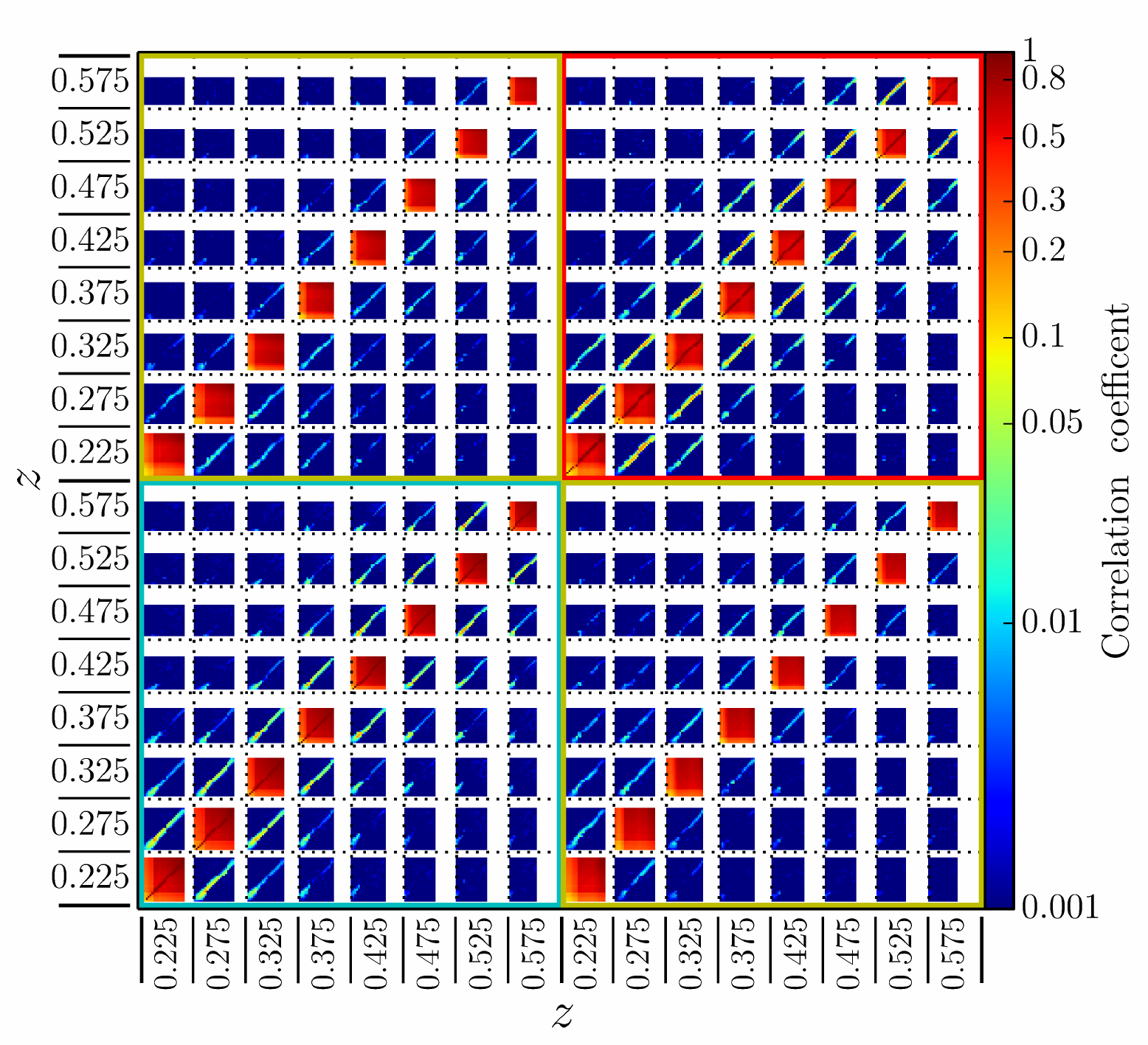}
     \caption{Correlation coefficients of the ALHAMBRA luminosity function covariance matrix, covering only the first eight redshift ranges for illustration purposes. The axes mark the redshift bin of interest for SF-SF galaxies (cyan square), Q-Q galaxies (red square), and SF-Q and Q-SF galaxies (green squares). Each redshift box delimited by dotted lines shows the $B-$band luminosity correlations. The cosmic variance highly correlates galaxies at the same redshift, with the correlations between different redshift and galaxy types caused by the photometric uncertainties.}
     \label{covmatrix}
\end{figure*}

The complete information about the correlations between redshift ranges, luminosities, and galaxy populations is encoded in the non-diagonal terms of the covariance matrix. For illustration purposes, we present the correlation coefficients of the first eight redshift bins under study in Fig.~\ref{covmatrix}. Several important features should be noted. The most relevant one is the impact of the cosmic variance term in the correlation between luminosities and galaxy types at a given redshift bin. The cosmic variance highly correlates both bright-faint and SF-Q galaxies, with correlations at the level of $\sim50$\%. This large correlation was anticipated by the theoretical work of \citet{smith12} and is confirmed by the ALHAMBRA cosmic variance measurements performed to compute the galaxy bias in Sect.~\ref{biasfunc}. We measured a correlation of $\sim40-60$\% at different luminosities and galaxy types, reinforcing therefore the estimated cosmic variance term.

The second important feature is the correlation between adjacent redshift and luminosity bins for galaxies of the same type (SF-SF and Q-Q). This correlation, at the level of $\sim10-20$\%, is due to the photometric redshift and absolute magnitude uncertainties. As expected, the correlation diminished as we compare well separated redshift bins.

The final remarkable feature is the correlation between quiescent and star-forming galaxies present in the shot noise term. This correlation, at the level of $\sim1-5$\%, was anticipated by the Fig.~\ref{pdfzt}. In the example presented in that figure, an early spectral galaxy solution is compatible with a late spectral galaxy solution located at higher redshifts. This redshift - galaxy type degeneracy emerges as the correlation of quiescent galaxies with those star-forming galaxies located at larger redshifts, and thus the Q-SF cross term is not symmetric in redshift. In addition, a fraction of the red population is counted as star-forming because of the dusty contamination detailed in Sect.~\ref{colpos}.

The covariance matrix of the luminosity function is fundamental to perform a realistic fitting to the data, accounting for all the possible correlations present in the observations. The studies in future large scale surveys, such as J-PAS, Euclid, or LSST, will benefit of the robust error budget encoded in the covariance matrix.

\section{Summary and conclusions}\label{conclusion}
We have studied the evolution of the $B$-band luminosity function since $z \sim 1$ both for star-forming and quiescent galaxies using ALHAMBRA data. We developed a novel methodology that statistically uses the output of current photometric redshift codes without losing information, reliably works with any pre-selection of the sources, neither in the $I$-band magnitude nor in colour, and provides an unbiased estimation of the luminosity function in multi-filter surveys.

We use the photometric redshift and the $I$-band selection magnitude PDFs of those ALHAMBRA galaxies with real magnitude $I_0 \leq 24$ to compute the posterior luminosity function $\Phi\,(z,M_B)$. We statistically study star-forming and quiescent galaxies thanks to the template information encoded in the PDFs. The luminosity function covariance matrix in redshift - magnitude - galaxy type space is computed, including the cosmic variance. That is estimated from the intrinsic dispersion of the luminosity function measurements in the 48 ALHAMBRA sub-fields. The uncertainty due to the photometric redshift prior is also included in our analysis. 

We modelled $\Phi\,(z,M_B)$ with a redshift-dependent Schechter function affected by the same selection effects than the data. The measured ALHAMBRA luminosity function at $0.2 \leq z < 1$ and the evolving Schechter parameters both for quiescent and star-forming galaxies agree with previous results in the literature.  The estimated redshift evolution of $M_{B}^* \propto Qz$ is $Q_{\rm SF} = -1.03 \pm 0.08$ and $Q_{\rm Q} = -0.80 \pm 0.08$, and of $\log_{10}\phi^{*} \propto Pz$ is $P_{\rm SF} = -0.01 \pm 0.03$ and $P_{\rm Q} = -0.41 \pm 0.05$. The measured faint-end slopes are $\alpha_{\rm SF} = -1.29 \pm 0.02$ and $\alpha_{\rm Q} = -0.53 \pm 0.04$. We find a significant population of faint quiescent galaxies with $M_B \gtrsim -18$, modelled by a second Schechter function with slope $\beta = -1.31 \pm 0.11$.

Our results implies a factor $2.55\pm0.14$ decrease in the luminosity density of the star-forming population since $z = 1$, reflecting the decrease of the star formation rate with time. We estimate a factor $1.25\pm0.16$ increase for the quiescent luminosity density, confirming the continuous build-up of the quiescent population since $z = 1$ to the present. The contribution of the faint quiescent population to the luminosity density increases from $3$\% at $z = 1$ to 6\% at $z = 0$.

The next generation large-area photometric surveys will benefit of the PROFUSE methodologies, and we are now ready to analyse with exquisite details the luminosity function of the J-PAS survey. Assuming a $I_0 \lesssim 22.5$ selection for the J-PAS galaxies, we will detect the faint quiescent up-turn up to $z \sim 0.3$, closing the gap between local Universe surveys and cosmological surveys. In addition, we will track the evolution of the bright population with overwhelming statistics up to $z \sim 1$, the redshift at which we cross the $M_B^{*}$ limit. The $B-$band luminosity function analysis presented in this paper will be expanded in the future to the $UV$ luminosity function at $z > 2.5$ and the stellar mass function at $z < 1$ with the ALHAMBRA $I-$band selected catalogue, and to the $K_{\rm s}$ luminosity function at $z < 2.5$ with the ALHAMBRA $K_{\rm s}-$band selected catalogue presented in \citet{nieves16}. Moreover, the emergence of the faint quiescent population has been related with environmental processes \citep[e.g.][]{peng10}, and we will explore also the impact of environment in the ALHAMBRA luminosity and stellar mass functions in future work.

\begin{acknowledgements}
We dedicate this paper to the memory of our six IAC colleagues and friends who
met with a fatal accident in Piedra de los Cochinos, Tenerife, in February 2007,
with special thanks to Maurizio Panniello, whose teachings of \texttt{python}
were so important for this paper. We thank R.~Angulo, S.~Bonoli, A.~Ederoclite, C.~Herna\'ndez-Monteagudo, A.~Mar\'{\i}n-Franch, A.~Orsi, and all the CEFCA staff, post-docs, and students for useful and productive discussions. We thank the anonymous referee for his/her suggestions.\\
This work has been mainly funded by the FITE (Fondos de Inversiones de Teruel) and the projects AYA2015-66211-C2-1, AYA2012-30789, AYA2006-14056, and CSD2007-00060. We also acknowledge support from the Spanish Ministry for Economy and Competitiveness and FEDER funds through grants AYA2010-15081, AYA2010-15169, AYA2010-22111-C03-01, AYA2010-22111-C03-02, AYA2011-29517-C03-01, AYA2012-39620, AYA2013-40611-P, AYA2013-42227-P, AYA2013-43188-P, AYA2013-48623-C2-1, AYA2013-48623-C2-2, ESP2013-48274, AYA2014-58861-C3-1, Arag\'on Government Research Group E103, Generalitat Valenciana projects Prometeo 2009/064 and PROMETEOII/2014/060, Junta de Andaluc\'{\i}a grants TIC114, JA2828, P10-FQM-6444, and Generalitat de Catalunya project SGR-1398. E.~T. acknowledges the support by the ETAg grants IUT26-2, IUT40-2, and by the European Regional Development Fund (TK133). A.~M. acknowledges the financial support of the Brazilian funding agency FAPESP (Post-doc fellowship - process number 2014/11806-9). B.~A. has received funding from the European Union's Horizon 2020 research and innovation programme under the Marie Sklodowska-Curie grant agreement No. 656354.
This research made use of \texttt{Astropy}, a community-developed core \texttt{Python} package for Astronomy \citep{astropy}, and \texttt{Matplotlib}, a 2D graphics package used for \texttt{Python} for publication-quality image generation across user interfaces and operating systems \citep{pylab}.
\end{acknowledgements}

\bibliography{biblio}
\bibliographystyle{aa}

\appendix

\begin{figure}[t]
\centering
\resizebox{\hsize}{!}{\includegraphics{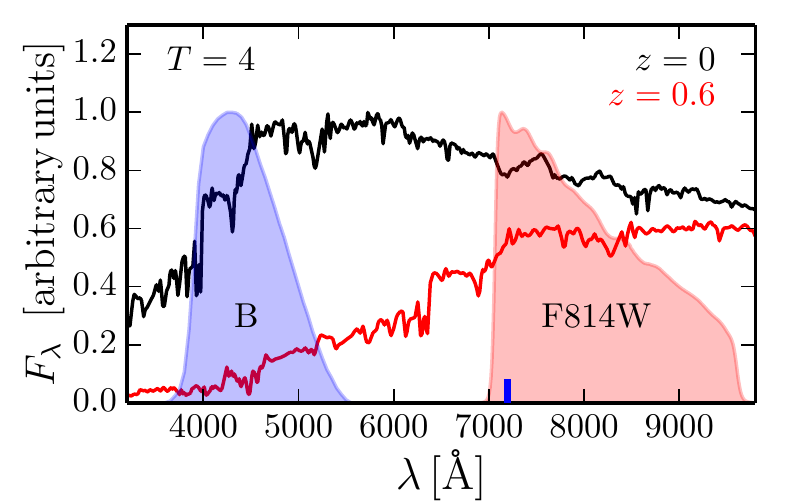}}
\caption{Example of $k$-correction for the \texttt{BPZ2} elliptical template $T = 4$. The rest-frame flux in arbitrary units (black line) is observed trough the $B$ filter (blue area), and the properly scaled and redshifted flux of the same template at $z = 0.6$ (red line) is observed trough the $I$ filter (red area). The blue tick marks the central wavelength of the $B$-band filter at the observed frame.}
\label{kcorrfig}
\end{figure}

\begin{figure}[t]
\centering
\resizebox{\hsize}{!}{\includegraphics{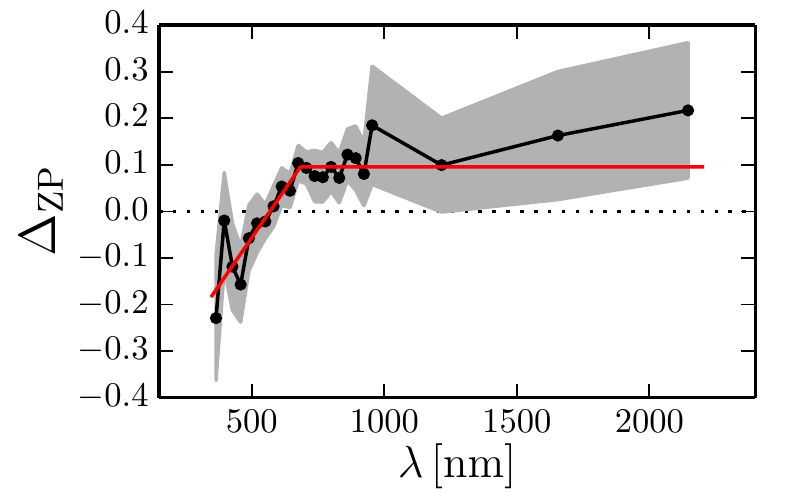}}\\
\resizebox{\hsize}{!}{\includegraphics{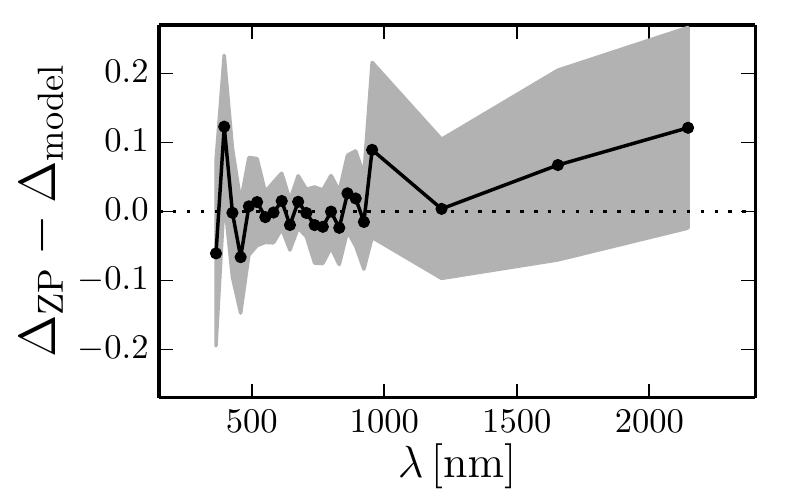}}
\caption{{\it Top panel} : Median zero point correction $\Delta_{\rm ZP}$ applied by \texttt{BPZ2} to the original ALHAMBRA photometry (black points). The observed structure is modelled with a linear function and a constant value (red solid line). {\it Bottom panel} : Residual zero point correction after removing the model for the median correction. The grey area in both panels shows the field-to-field dispersion of the correction.}
\label{zpoff}
\end{figure}

\section{$k-$correction with \texttt{BPZ2}}\label{kcorr}
The estimation of the $B-$band absolute magnitude described in Sect.~\ref{zmbpos} relies on the $k$-correction between the observed $I$ band of the source at redshift $z$ and the targeted $B$ band of the source at rest-frame, as illustrated in Fig.~\ref{kcorrfig}. We computed such correction for each $z$ and \texttt{BPZ2} template, the two variables covered by the photometric redshift PDFs. The observed $I$ passband, noted ${\rm R}_{I}$, corresponds to the HST/$F814W$ filter, and the targeted $B$ passband, noted ${\rm R}_{B}$, is a standard $B$ Johnson filter. To facilitate the comparison with future work, both assumed filter curves are available at the PROFUSE web page.

The $k-$correction as a function of $z$ and $T$ in Eq.~(\ref{mbzteq}) is defined as
\begin{align}
k\,(z,T) = &\ {\rm ZP}\,(z) - 2.5\log_{10}\Bigg[\frac{1}{1+z} \frac{\int \lambda\,F_{\lambda}^T\big(\frac{\lambda}{1+z}\big)\,{\rm R}_I\,{\rm d}\lambda}{\int \lambda\,{\rm R}_I\,{\rm d}\lambda}\Bigg] + \nonumber\\
& 2.5\log_{10}\Bigg[\frac{\int \lambda\,F_{\lambda}^T(\lambda)\,{\rm R}_B\,{\rm d}\lambda}{\int \lambda\,{\rm R}_B\,{\rm d}\lambda}\Bigg] - 1.29,\label{kcorreq}
\end{align}
where the constant term accounts for the $\lambda$ to $\nu$ (i.e. AB magnitude) transformation, $F_{\lambda}^T$ is the rest-frame flux of the \texttt{BPZ2} template $T$, and ${\rm ZP}\,(z)$ is the zero point colour term applied by \texttt{BPZ2} to the data. As other photometric redshift codes, a zero point recalibration is performed by \texttt{BPZ2} at each ALHAMBRA sub-field to improve the accuracy of the photometric redshifts \citep[see][for details]{molino13}. We noted that the median zero point correction in ALHAMBRA presents a structure (Fig.~\ref{zpoff}) that translates to a colour term at $z \lesssim 0.5$. This colour term makes low redshift galaxies bluer than expected, and therefore brighter in the rest-frame $B$ band. We described this structure with a linear plus constant model, and derived the zero point colour term ${\rm ZP}\,(z)$ in Eq.~\ref{kcorreq} from it. This colour term is zero at $z > 0.5$, and decreases linearly down to ${\rm ZP}\,(0) = -0.19$ at $z < 0.5$. We removed our model from the original zero point correction, revealing the expected filter-to-filter refinements in the photometric zero points (Fig.~\ref{zpoff}). We stress that the same ${\rm ZP}\,(z)$ was applied to all the ALHAMBRA sub-fields, so the filter-to-filter zero point correction is preserved.

Finally, we computed the magnitude difference between our fiducial $B-$band filter and those assumed in the literature. Only two studies do not target a standard $B$ Johnson filter as reference: \citet{beare15} assume a $B$ Bessel filter, and \citet{loveday12} a $g$ SDSS filter at $z = 0.1$. In the first case, the expected difference is bellow 0.02 mag for any template, so we applied no offset to the values provided by \citet{beare15}. In the second case, the expected difference for blue galaxies ($T = {\rm S/SB}$) is lower than 0.02 mag, but reaches 0.05 mag for red galaxies ($T = {\rm E/S0}$). Thus, we applied a -0.05 mag offset to the magnitudes of red galaxies provided by \citet{loveday12}.

\section{ALHAMBRA luminosity and galaxy bias function values}\label{tables}
In this Appendix we present the Tables with the ALHAMBRA luminosity ($\Delta z = 0.2$) and galaxy bias functions of both star-forming (Tables~\ref{lfblue_tab} and \ref{biasblue_tab}) and quiescent (Tables~\ref{lfred_tab} and \ref{biasred_tab}) galaxies. The luminosity function computed with $\Delta z = 0.05, 0.1, 0.2$ and their covariance matrices are accessible at the PROFUSE web page. Those bins with $V_{\rm eff}/\Delta V < 0.01$ were discarded.

\begin{table*}
\caption{ALHAMBRA luminosity function of star-forming galaxies $\Phi\,(z,M_B\,|\,{\rm SF})$.}
\label{lfblue_tab}
\begin{center}
\begin{tabular}{@{\extracolsep{2pt}}cccccccccc@{}}
\hline\hline\noalign{\smallskip}
$M_B^{-}$ & $M_B^{+}$ & \multicolumn{2}{c}{$0.2 \leq z < 0.4$} & \multicolumn{2}{c}{$0.4 \leq z < 0.6$} & \multicolumn{2}{c}{$0.6 \leq z < 0.8$} & \multicolumn{2}{c}{$0.8 \leq z < 1.0$}\\\noalign{\smallskip}\cline{3-4}\cline{5-6}\cline{7-8}\cline{9-10}\noalign{\smallskip}

 & & $\langle M_B \rangle$ & $\log_{10}\Phi$ & $\langle M_B \rangle$ & $\log_{10}\Phi$ & $\langle M_B \rangle$ & $\log_{10}\Phi$ & $\langle M_B \rangle$ & $\log_{10}\Phi$\\
\noalign{\smallskip}
\hline
\noalign{\smallskip}
$-24.5$ & $-23.0$ & $-23.29$ & $-7.05 \pm 0.34$ & $-23.42$ & $-6.39 \pm 0.26$ & $-23.20$ & $-5.75 \pm 0.14$ & $-23.34$ & $-5.45 \pm 0.09$\\
$-23.0$ & $-22.7$ & $-22.79$ & $-5.56 \pm 0.21$ & $-22.81$ & $-5.02 \pm 0.11$ & $-22.82$ & $-4.64 \pm 0.06$ & $-22.83$ & $-4.44 \pm 0.05$\\
$-22.7$ & $-22.4$ & $-22.56$ & $-5.19 \pm 0.17$ & $-22.54$ & $-4.55 \pm 0.07$ & $-22.52$ & $-4.18 \pm 0.05$ & $-22.53$ & $-3.99 \pm 0.04$\\
$-22.4$ & $-22.1$ & $-22.21$ & $-4.66 \pm 0.11$ & $-22.22$ & $-4.14 \pm 0.06$ & $-22.23$ & $-3.83 \pm 0.04$ & $-22.23$ & $-3.59 \pm 0.03$\\
$-22.1$ & $-21.8$ & $-21.91$ & $-4.13 \pm 0.06$ & $-21.93$ & $-3.70 \pm 0.04$ & $-21.93$ & $-3.51 \pm 0.03$ & $-21.94$ & $-3.27 \pm 0.02$\\
$-21.8$ & $-21.5$ & $-21.62$ & $-3.63 \pm 0.05$ & $-21.63$ & $-3.38 \pm 0.03$ & $-21.64$ & $-3.23 \pm 0.02$ & $-21.64$ & $-3.08 \pm 0.02$\\
$-21.5$ & $-21.2$ & $-21.34$ & $-3.30 \pm 0.04$ & $-21.34$ & $-3.15 \pm 0.03$ & $-21.34$ & $-3.03 \pm 0.02$ & $-21.34$ & $-2.90 \pm 0.02$\\
$-21.2$ & $-20.9$ & $-21.04$ & $-3.10 \pm 0.03$ & $-21.04$ & $-2.99 \pm 0.02$ & $-21.04$ & $-2.87 \pm 0.02$ & $-21.04$ & $-2.76 \pm 0.01$\\
$-20.9$ & $-20.6$ & $-20.74$ & $-2.94 \pm 0.03$ & $-20.74$ & $-2.85 \pm 0.02$ & $-20.74$ & $-2.75 \pm 0.02$ & $-20.75$ & $-2.64 \pm 0.01$\\
$-20.6$ & $-20.3$ & $-20.44$ & $-2.76 \pm 0.03$ & $-20.45$ & $-2.74 \pm 0.02$ & $-20.45$ & $-2.65 \pm 0.01$ & $-20.45$ & $-2.56 \pm 0.01$\\
$-20.3$ & $-20.0$ & $-20.15$ & $-2.65 \pm 0.02$ & $-20.15$ & $-2.65 \pm 0.02$ & $-20.15$ & $-2.58 \pm 0.01$ & $-20.15$ & $-2.47 \pm 0.01$\\
$-20.0$ & $-19.7$ & $-19.85$ & $-2.56 \pm 0.02$ & $-19.84$ & $-2.58 \pm 0.02$ & $-19.85$ & $-2.50 \pm 0.01$ & $-19.85$ & $-2.42 \pm 0.01$\\
$-19.7$ & $-19.4$ & $-19.55$ & $-2.48 \pm 0.02$ & $-19.55$ & $-2.50 \pm 0.02$ & $-19.55$ & $-2.45 \pm 0.01$ & $-19.56$ & $-2.36 \pm 0.01$\\
$-19.4$ & $-19.1$ & $-19.25$ & $-2.43 \pm 0.02$ & $-19.25$ & $-2.45 \pm 0.01$ & $-19.25$ & $-2.41 \pm 0.01$ & $-19.27$ & $-2.34 \pm 0.02$\\
$-19.1$ & $-18.8$ & $-18.95$ & $-2.38 \pm 0.02$ & $-18.95$ & $-2.39 \pm 0.01$ & $-18.95$ & $-2.37 \pm 0.01$ & $-19.02$ & $-2.30 \pm 0.03$\\
$-18.8$ & $-18.5$ & $-18.65$ & $-2.32 \pm 0.02$ & $-18.65$ & $-2.33 \pm 0.01$ & $-18.67$ & $-2.37 \pm 0.01$ & $\cdots$ & $\cdots$\\
$-18.5$ & $-18.2$ & $-18.35$ & $-2.27 \pm 0.02$ & $-18.35$ & $-2.29 \pm 0.01$ & $-18.42$ & $-2.34 \pm 0.03$ & $\cdots$ & $\cdots$\\
$-18.2$ & $-17.9$ & $-18.05$ & $-2.22 \pm 0.02$ & $-18.06$ & $-2.23 \pm 0.02$ & $\cdots$ & $\cdots$ & $\cdots$ & $\cdots$\\
$-17.9$ & $-17.6$ & $-17.75$ & $-2.17 \pm 0.02$ & $-17.76$ & $-2.18 \pm 0.02$ & $\cdots$ & $\cdots$ & $\cdots$ & $\cdots$\\
$-17.6$ & $-17.3$ & $-17.45$ & $-2.12 \pm 0.01$ & $-17.49$ & $-2.17 \pm 0.03$ & $\cdots$ & $\cdots$ & $\cdots$ & $\cdots$\\
$-17.3$ & $-17.0$ & $-17.16$ & $-2.06 \pm 0.02$ & $\cdots$ & $\cdots$ & $\cdots$ & $\cdots$ & $\cdots$ & $\cdots$\\
$-17.0$ & $-16.7$ & $-16.86$ & $-2.01 \pm 0.02$ & $\cdots$ & $\cdots$ & $\cdots$ & $\cdots$ & $\cdots$ & $\cdots$\\
$-16.7$ & $-16.4$ & $-16.56$ & $-1.98 \pm 0.02$ & $\cdots$ & $\cdots$ & $\cdots$ & $\cdots$ & $\cdots$ & $\cdots$\\
$-16.4$ & $-16.1$ & $-16.26$ & $-1.92 \pm 0.03$ & $\cdots$ & $\cdots$ & $\cdots$ & $\cdots$ & $\cdots$ & $\cdots$\\
$-16.1$ & $-15.8$ & $-15.97$ & $-1.87 \pm 0.04$ & $\cdots$ & $\cdots$ & $\cdots$ & $\cdots$ & $\cdots$ & $\cdots$\\
$-15.8$ & $-15.5$ & $-15.69$ & $-1.83 \pm 0.07$ & $\cdots$ & $\cdots$ & $\cdots$ & $\cdots$ & $\cdots$ & $\cdots$\\
\hline
\end{tabular}
\end{center}
\tablefoot{The units of the luminosity function are Mpc$^{-3}$\,mag$^{-1}$. The quoted uncertainties only reflect the diagonal terms of the covariance matrix $\boldsymbol{\Sigma}_{\Phi}$, both shot noise and cosmic variance.
}
\end{table*}

\begin{table*}
\caption{ALHAMBRA galaxy bias function of star-forming galaxies $b_v\,(z,M_B\,|\,{\rm SF})$.}
\label{biasblue_tab}
\begin{center}
\begin{tabular}{@{\extracolsep{2pt}}ccccccc@{}}
\hline\hline\noalign{\smallskip}
Luminosity range & \multicolumn{2}{c}{$0.2 \leq z < 0.65$} & \multicolumn{2}{c}{$0.65 \leq z < 0.85$} & \multicolumn{2}{c}{$0.85 \leq z < 1$}\\\noalign{\smallskip}\cline{2-3}\cline{4-5}\cline{6-7}\noalign{\smallskip}

 & $\langle M_B - M_B^* \rangle$ & $b_v$ & $\langle  M_B - M_B^* \rangle$ & $b_v$ & $\langle  M_B - M_B^* \rangle$ & $b_v$\\
\noalign{\smallskip}
\hline
\noalign{\smallskip}
$M_B - M_B^* \leq 0$        & $-0.42$  & $1.62 \pm 0.26$ & $-0.44$   & $1.77 \pm 0.25$ & $-0.42$ & $1.49 \pm 0.24$\\
$0 < M_B - M_B^* \leq 0.5$   & $0.27$  & $1.22 \pm 0.21$ &  $0.27$   & $1.28 \pm 0.21$ & $0.27$  & $1.37 \pm 0.21$\\
$0.5 < M_B - M_B^* \leq 1.0$ & $0.76$  & $1.18 \pm 0.19$ &  $0.76$   & $1.19 \pm 0.18$ & $0.76$  & $1.26 \pm 0.16$\\
$1.0 < M_B - M_B^* \leq 1.5$ & $1.26$  & $1.12 \pm 0.18$ &  $1.26$   & $0.98 \pm 0.14$ & $1.26$  & $1.15 \pm 0.15$\\
$1.5 < M_B - M_B^* \leq 2.0$ & $1.76$  & $0.94 \pm 0.15$ &  $1.76$   & $0.95 \pm 0.14$ & $\cdots$ & $\cdots$\\
$2.0 < M_B - M_B^* \leq 2.5$ & $2.26$  & $0.78 \pm 0.13$ &  $\cdots$ & $\cdots$ & $\cdots$ & $\cdots$\\
\hline
\end{tabular}
\end{center}
\end{table*}

\begin{table*}
\caption{ALHAMBRA luminosity function of quiescent galaxies $\Phi\,(z,M_B\,|\,{\rm Q})$.}
\label{lfred_tab}
\begin{center}
\begin{tabular}{@{\extracolsep{2pt}}cccccccccc@{}}
\hline\hline\noalign{\smallskip}
$M_B^{-}$ & $M_B^{+}$ & \multicolumn{2}{c}{$0.2 \leq z < 0.4$} & \multicolumn{2}{c}{$0.4 \leq z < 0.6$} & \multicolumn{2}{c}{$0.6 \leq z < 0.8$} & \multicolumn{2}{c}{$0.8 \leq z < 1.0$}\\\noalign{\smallskip}\cline{3-4}\cline{5-6}\cline{7-8}\cline{9-10}\noalign{\smallskip}
 & & $\langle M_B \rangle$ & $\log_{10}\Phi$ & $\langle M_B \rangle$ & $\log_{10}\Phi$ & $\langle M_B \rangle$ & $\log_{10}\Phi$ & $\langle M_B \rangle$ & $\log_{10}\Phi$\\
\noalign{\smallskip}
\hline
\noalign{\smallskip}
$-24.5$ & $-23.0$ & $-23.07$ & $-6.75 \pm 0.38$ & $-23.37$ & $-6.37 \pm 0.33$ & $-23.13$ & $-6.29 \pm 0.21$ & $-23.18$ & $-5.83 \pm 0.13$\\
$-23.0$ & $-22.7$ & $-22.83$ & $-5.05 \pm 0.29$ & $-22.80$ & $-5.08 \pm 0.20$ & $-22.80$ & $-4.73 \pm 0.09$ & $-22.83$ & $-4.68 \pm 0.06$\\
$-22.7$ & $-22.4$ & $-22.53$ & $-4.87 \pm 0.21$ & $-22.52$ & $-4.44 \pm 0.09$ & $-22.53$ & $-4.18 \pm 0.06$ & $-22.53$ & $-4.14 \pm 0.05$\\
$-22.4$ & $-22.1$ & $-22.21$ & $-4.20 \pm 0.10$ & $-22.23$ & $-4.05 \pm 0.06$ & $-22.23$ & $-3.90 \pm 0.05$ & $-22.23$ & $-3.82 \pm 0.04$\\
$-22.1$ & $-21.8$ & $-21.94$ & $-3.79 \pm 0.06$ & $-21.94$ & $-3.76 \pm 0.05$ & $-21.94$ & $-3.68 \pm 0.04$ & $-21.94$ & $-3.61 \pm 0.03$\\
$-21.8$ & $-21.5$ & $-21.63$ & $-3.53 \pm 0.05$ & $-21.64$ & $-3.52 \pm 0.04$ & $-21.64$ & $-3.51 \pm 0.03$ & $-21.65$ & $-3.49 \pm 0.02$\\
$-21.5$ & $-21.2$ & $-21.34$ & $-3.30 \pm 0.04$ & $-21.34$ & $-3.37 \pm 0.03$ & $-21.34$ & $-3.37 \pm 0.03$ & $-21.34$ & $-3.39 \pm 0.02$\\
$-21.2$ & $-20.9$ & $-21.05$ & $-3.15 \pm 0.04$ & $-21.05$ & $-3.26 \pm 0.03$ & $-21.05$ & $-3.28 \pm 0.02$ & $-21.05$ & $-3.31 \pm 0.02$\\
$-20.9$ & $-20.6$ & $-20.74$ & $-3.09 \pm 0.03$ & $-20.74$ & $-3.20 \pm 0.03$ & $-20.75$ & $-3.26 \pm 0.02$ & $-20.75$ & $-3.31 \pm 0.02$\\
$-20.6$ & $-20.3$ & $-20.44$ & $-2.99 \pm 0.03$ & $-20.45$ & $-3.21 \pm 0.03$ & $-20.45$ & $-3.26 \pm 0.02$ & $-20.45$ & $-3.29 \pm 0.02$\\
$-20.3$ & $-20.0$ & $-20.15$ & $-2.96 \pm 0.03$ & $-20.15$ & $-3.23 \pm 0.02$ & $-20.15$ & $-3.19 \pm 0.02$ & $-20.15$ & $-3.28 \pm 0.02$\\
$-20.0$ & $-19.7$ & $-19.85$ & $-2.95 \pm 0.03$ & $-19.85$ & $-3.24 \pm 0.02$ & $-19.85$ & $-3.20 \pm 0.02$ & $-19.86$ & $-3.32 \pm 0.02$\\
$-19.7$ & $-19.4$ & $-19.55$ & $-2.97 \pm 0.03$ & $-19.55$ & $-3.30 \pm 0.02$ & $-19.55$ & $-3.21 \pm 0.02$ & $-19.56$ & $-3.33 \pm 0.02$\\
$-19.4$ & $-19.1$ & $-19.25$ & $-3.07 \pm 0.03$ & $-19.25$ & $-3.33 \pm 0.02$ & $-19.25$ & $-3.23 \pm 0.02$ & $-19.27$ & $-3.32 \pm 0.02$\\
$-19.1$ & $-18.8$ & $-18.95$ & $-3.12 \pm 0.03$ & $-18.95$ & $-3.34 \pm 0.02$ & $-18.95$ & $-3.27 \pm 0.02$ & $-19.02$ & $-3.26 \pm 0.04$\\
$-18.8$ & $-18.5$ & $-18.65$ & $-3.18 \pm 0.03$ & $-18.65$ & $-3.34 \pm 0.02$ & $-18.67$ & $-3.31 \pm 0.02$ & $\cdots$ & $\cdots$\\
$-18.5$ & $-18.2$ & $-18.35$ & $-3.20 \pm 0.03$ & $-18.35$ & $-3.35 \pm 0.02$ & $-18.42$ & $-3.26 \pm 0.04$ & $\cdots$ & $\cdots$\\
$-18.2$ & $-17.9$ & $-18.05$ & $-3.23 \pm 0.02$ & $-18.06$ & $-3.34 \pm 0.02$ & $\cdots$ & $\cdots$ & $\cdots$ & $\cdots$\\
$-17.9$ & $-17.6$ & $-17.74$ & $-3.16 \pm 0.02$ & $-17.76$ & $-3.29 \pm 0.03$ & $\cdots$ & $\cdots$ & $\cdots$ & $\cdots$\\
$-17.6$ & $-17.3$ & $-17.45$ & $-3.06 \pm 0.02$ & $-17.48$ & $-3.22 \pm 0.04$ & $\cdots$ & $\cdots$ & $\cdots$ & $\cdots$\\
$-17.3$ & $-17.0$ & $-17.15$ & $-2.98 \pm 0.02$ & $\cdots$ & $\cdots$ & $\cdots$ & $\cdots$ & $\cdots$ & $\cdots$\\
$-17.0$ & $-16.7$ & $-16.86$ & $-2.90 \pm 0.03$ & $\cdots$ & $\cdots$ & $\cdots$ & $\cdots$ & $\cdots$ & $\cdots$\\
$-16.7$ & $-16.4$ & $-16.56$ & $-2.82 \pm 0.03$ & $\cdots$ & $\cdots$ & $\cdots$ & $\cdots$ & $\cdots$ & $\cdots$\\
$-16.4$ & $-16.1$ & $-16.26$ & $-2.69 \pm 0.04$ & $\cdots$ & $\cdots$ & $\cdots$ & $\cdots$ & $\cdots$ & $\cdots$\\
$-16.1$ & $-15.8$ & $-15.97$ & $-2.58 \pm 0.05$ & $\cdots$ & $\cdots$ & $\cdots$ & $\cdots$ & $\cdots$ & $\cdots$\\
$-15.8$ & $-15.5$ & $-15.69$ & $-2.52 \pm 0.09$ & $\cdots$ & $\cdots$ & $\cdots$ & $\cdots$ & $\cdots$ & $\cdots$\\
\hline
\end{tabular}
\end{center}
\tablefoot{The units of the luminosity function are Mpc$^{-3}$\,mag$^{-1}$. The quoted uncertainties only reflect the diagonal terms of the covariance matrix $\boldsymbol{\Sigma}_{\Phi}$, both shot noise and cosmic variance.
}
\end{table*}

\begin{table*}
\caption{ALHAMBRA galaxy bias function of quiescent galaxies $b_v\,(z,M_B\,|\,{\rm Q})$.}
\label{biasred_tab}
\begin{center}
\begin{tabular}{@{\extracolsep{2pt}}ccccccc@{}}
\hline\hline\noalign{\smallskip}
Luminosity range & \multicolumn{2}{c}{$0.2 \leq z < 0.65$} & \multicolumn{2}{c}{$0.65 \leq z < 0.85$} & \multicolumn{2}{c}{$0.85 \leq z < 1$}\\\noalign{\smallskip}\cline{2-3}\cline{4-5}\cline{6-7}\noalign{\smallskip}

 & $\langle M_B - M_B^* \rangle$ & $b_v$ & $\langle  M_B - M_B^* \rangle$ & $b_v$ & $\langle  M_B - M_B^* \rangle$ & $b_v$\\
\noalign{\smallskip}
\hline
\noalign{\smallskip}
$M_B - M_B^* \leq 0$         & $-0.56$ & $1.53 \pm 0.27$ & $-0.57$  & $2.63 \pm 0.38$ & $-0.55$  & $2.12 \pm 0.31$\\
$0 < M_B - M_B^* \leq 0.5$   & $0.25$  & $1.41 \pm 0.28$ & $0.25$   & $1.76 \pm 0.29$ & $0.25$   & $1.65 \pm 0.25$\\
$0.5 < M_B - M_B^* \leq 1.0$ & $0.75$  & $1.54 \pm 0.31$ & $0.76$   & $1.31 \pm 0.25$ & $0.75$   & $1.11 \pm 0.34$\\
$1.0 < M_B - M_B^* \leq 1.5$ & $1.24$  & $0.88 \pm 0.43$ & $1.25$   & $1.25 \pm 0.27$ & $\cdots$ & $\cdots$\\
$1.5 < M_B - M_B^* \leq 2.0$ & $1.75$  & $1.02 \pm 0.42$ & $1.74$   & $1.38 \pm 0.30$ & $\cdots$ & $\cdots$\\
$2.0 < M_B - M_B^* \leq 2.5$ & $2.25$  & $1.37 \pm 0.33$ & $\cdots$ & $\cdots$        & $\cdots$ & $\cdots$\\
\hline
\end{tabular}
\end{center}
\end{table*}

\end{document}